\documentclass[acmsmall,screen]{acmart}
\AtBeginDocument{%
  }

\acmISBN{978-1-4503-XXXX-X/2018/06}

\usepackage{enumitem}
\usepackage{tabularx}
\usepackage{pifont}
\usepackage{multirow}
\usepackage{graphicx}
\usepackage{hhline}
\usepackage{booktabs}
\usepackage[table]{xcolor}
\usepackage{subcaption}
\usepackage{calc}
\usepackage{float}
\usepackage[framemethod=TikZ]{mdframed}
\usepackage{multicol}   
\usepackage{enumitem}   
\usepackage{array}
\usepackage[normalem]{ulem}

\usepackage{algorithm}
\usepackage{algpseudocode}

\newmdenv[
  topline=false,
  bottomline=false,
  rightline=false,
  linewidth=3pt,          
  linecolor=gray!60,      
  backgroundcolor=gray!10, 
  innerleftmargin=10pt,
  innerrightmargin=10pt,
  innertopmargin=8pt,
  innerbottommargin=8pt,
  skipabove=3pt,
  skipbelow=3pt
]{summarybox}

\newcommand{\techname}{\textsc{LoBREST}}
\newcommand{\restler}{\textsc{Restler}}
\newcommand{\evo}{\textsc{Evo-BB}}
\newcommand{\evomaster}{\textsc{EvoMaster}}
\newcommand{\rtg}{\textsc{RestTestGen}}
\newcommand{\morest}{\textsc{Morest}}
\newcommand{\restct}{\textsc{Restct}}
\newcommand{\schemathesis}{\textsc{Schemathesis}}
\newcommand{\arat}{\textsc{Arat-rl}}
\newcommand{\deeprest}{\textsc{Deeprest}}

\begin{document}

\title{Log-based, Business-aware REST API Testing}

\author{Ding Yang}
\authornote{Both authors contributed equally to this research.}
\email{dingyang@smail.nju.edu.cn}
\author{Ruixiang Qian}
\authornotemark[1]
\email{qianrx@smail.nju.edu.cn}
\affiliation{%
  \institution{Nanjing University}
  \city{Nanjing}
  \country{China}
}

\author{Zhao Wei}
\affiliation{
  \institution{Tencent}
  \country{China}
  }
\email{zachwei@tencent.com}

\author{Zhenyu Chen}
\affiliation{%
  \institution{Nanjing University}
  \country{China}
}
\email{zychen@nju.edu.cn}

\author{Chunrong Fang}
\affiliation{%
  \institution{Nanjing University}
  \country{China}
}
\email{fangchunrong@nju.edu.cn}







\begin{abstract}
  REST APIs enable collaboration among microservices.
  A single fault in a REST API can bring down the entire microservice system and cause significant financial losses, underscoring the importance of REST API testing.
  Effectively testing REST APIs requires thoroughly exercising the functionalities behind them.
  To this end, existing techniques leverage REST specifications (e.g., Swagger or OpenAPI) to generate test cases.
  Using the resource constraints extracted from specifications, these techniques work well for testing simple, business-insensitive functionalities, such as resource creation, retrieval, update, and deletion.
  However, for complex, business-sensitive functionalities, these specification-based techniques often fall short,
  since exercising such functionalities requires additional business constraints that are typically absent from REST specifications.

  In this paper, we present \techname{}, a log-based, business-aware REST API testing technique that leverages historical request logs (HRLogs) to effectively exercise the business-sensitive functionalities behind REST APIs.
  To obtain compact operation sequences that preserve clean and complete business constraints, \techname{} first employs a locality-slicing strategy to partition HRLogs into smaller slices.
  Then, to ensure the effectiveness of the obtained slices, \techname{} enhances them in two steps: (1) adding slices for operations missing from HRLogs, and (2) completing missing resources within the slices.
  Finally, to improve test adequacy, \techname{} uses these enhanced slices as initial seeds to perform business-aware fuzzing.
  We evaluate \techname{} on 17 real-world REST services and compare it with eight existing REST API testing tools, including the state-of-the-art tools \arat{}, \morest{}, and \deeprest{}.
  The experimental results demonstrate that \techname{} achieves the highest operation coverage on 16 services and the highest line coverage on 15 services.
  Specifically, \techname{} covers $2.1\times$ as many operations and $1.2\times$ as many lines as the second-best tool on average.
  Moreover, across all 17 services, \techname{} detects the most \texttt{5XX} bugs---108 in total, including 38 bugs that other tools fail to find.
\end{abstract}

\ccsdesc[500]{Software and its engineering~Software testing and debugging}

\keywords{REST API Testing, Fuzzing, Log Analysis, Business}


\maketitle

\section{Introduction} \label{sec:intro}
Web APIs constitute fundamental infrastructures for communication among distributed software systems \cite{newman2021building, pautasso2008restful}.
As an API design paradigm, Representational State Transfer (REST) has achieved widespread adoption due to its simplicity and flexibility \cite{masse2011rest}.
Today, 93\% of API services are constructed using the REST style, commonly referred to as REST Services \cite{api2025report}.
In large enterprises such as Amazon \cite{amazon} and Google \cite{google}, REST services expose a wide range of APIs and process billions of requests daily \cite{microservice2025report}.
A single bug in any of these APIs can propagate through services and bring down the entire system, highlighting the importance of REST API testing.

\begin{figure}
    \centering
    \includegraphics[width=\linewidth]{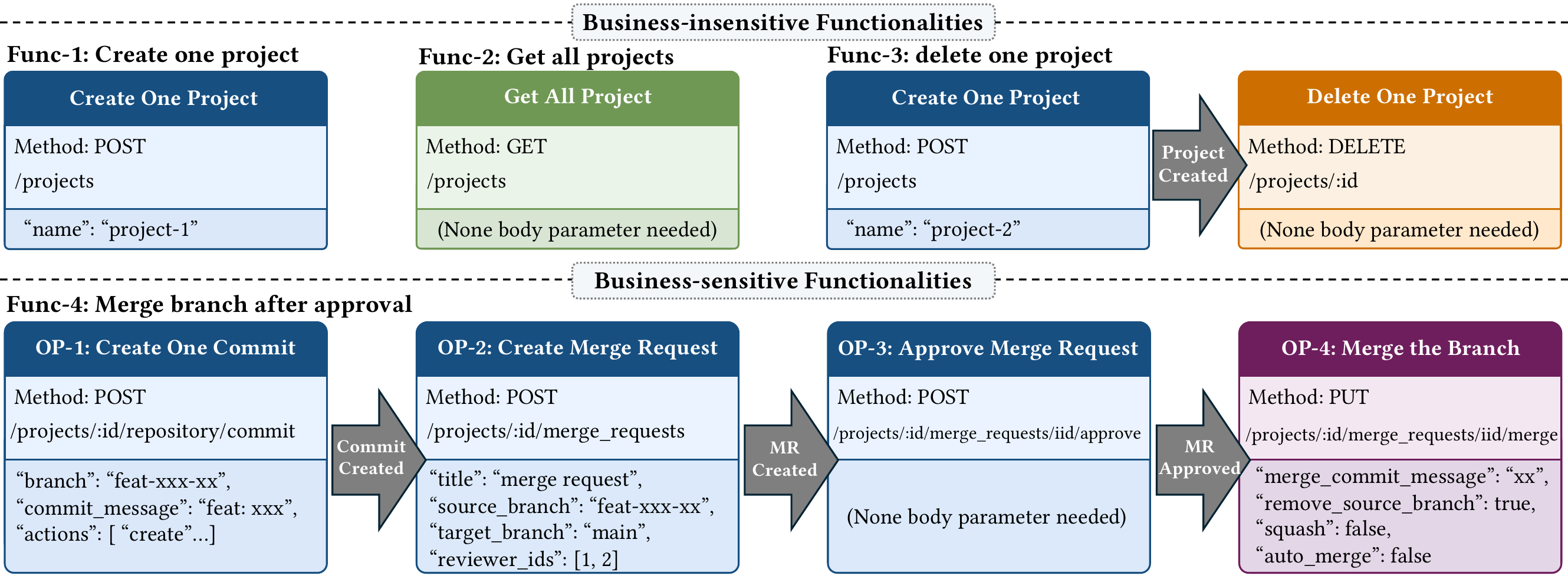}
    \caption{Examples of business-insensitive and business-sensitive functionalities in the GitLab REST service.}
    \label{fig:business-functionality-example}
    \Description{}
\end{figure}
Thorough testing of REST APIs requires comprehensively exercising the underlying functionalities, each of which typically involves multiple API operations.
These functionalities can be classified into two categories: \emph{business-insensitive} and \emph{business-sensitive}.
A business is relevant to the application domain of a REST service.
For instance, the business of an e-commerce service is shopping, and the business of a GitLab\footnote{GitLab is a popular DevOps platform serving over 10 million users, providing version control, CI/CD, and project management for software development and operation teams.} service is software development.
\autoref{fig:business-functionality-example} presents examples of the two types of functionalities in a GitLab REST service.
As shown in \autoref{fig:business-functionality-example}, a \textit{business-insensitive} functionality (such as Func-1,2,3 in \autoref{fig:business-functionality-example}) corresponds to basic resource management behaviors, i.e., creation, retrieval, update, and deletion (CRUD).
The execution of a business-insensitive functionality is generally trivial, requiring straightforward combinations of a few operations over the resources involved.
In contrast, \textit{business-sensitive functionalities} involve operations that are closely tied to the core service business; executing such functionalities requires not only invoking basic CRUD operations (OP-1 and OP-2) but also performing business-specific actions (OP-3 and OP-4) on the created resources, which is more challenging.
For example, Func-4 in \autoref{fig:business-functionality-example} is a business-sensitive functionality of GitLab that implements the action chain to ``merge branch after approval''.
To perform Func-4, an approval operation must be executed shortly after a merge request is created; otherwise, the merge operation will fail and the functionality will \textit{not} be fully tested.

To ensure the reliability of REST APIs, various testing techniques have been proposed \cite{atlidakis2019restler, hatfield2022schemathesis, karlsson2020quickrest, arcuri2019evomaster, viglianisi2020resttestgen, kim2022automated, liu2022morest, wu2022restct, corradini2024deeprest, kim2025llamaresttest}.
Generally, these techniques generate test cases as sequences of API operations based on Swagger \cite{swagger}/OpenAPI specifications \cite{openapi}, synthesizing operation sequences under \emph{resource constraints}.
For example, if operation A consumes a resource created by operation B, then there is a resource constraint that requires A to be executed after B.
With resource constraints, existing techniques usually work well in testing business-insensitive functionalities; however, they fall short when applied to business-sensitive functionalities.
The reason is that correctly exercising business-sensitive functionalities requires not only satisfying resource constraints but also adherence to \emph{business constraints}, which are typically \textit{not} readily available in REST specifications.
For example, fully exercising Func-4 in \autoref{fig:business-functionality-example} requires satisfying the following business constraints: (1) a strict execution order (from OP-1 to OP-4), (2) valid parameter combinations (e.g., \texttt{source\_branch} and \texttt{target\_branch} must exist simultaneously in OP-2), and (3) valid parameter values (e.g., parameter \texttt{action} in OP-1 can only take predefined values like \texttt{create} or \texttt{delete}).
Being unaware of business constraints, existing techniques fail to construct complete and valid operation sequences for business-sensitive functionalities.

In this paper, we propose \techname{}, a log-based, business-aware REST API testing technique that leverages historical request logs (HRLogs) to facilitate testing of business-sensitive functionalities in REST services.
Our key insight is that HRLogs inherently capture the business constraints needed for testing business-sensitive functionalities.
This is because, when users invoke business-sensitive functionalities, HRLogs will record the executed API operations as temporally ordered request traces, which implicitly encode both the execution-order and parameter-usage constraints.
By exploiting HRLogs, \techname{} can generate operation sequences that preserve business constraints, thereby enabling more effective testing and deeper exploration of these business-sensitive functionalities.

Despite the potential of HRLogs, effectively repurposing them for REST API testing is still non-trivial and poses three challenges:

\begin{itemize}[leftmargin=10pt]

\item \textbf{Challenge-1: How to preserve clean and complete business constraints?}
HRLogs are chaotic, containing requests from multiple users at different times.
Operations in a single functionality are usually fragmented across the entire logs.
Without filtering irrelevant operations and reassembling the relevant ones, the generated operation sequences still cannot successfully exercise the target business-sensitive functionality.

\item \textbf{Challenge-2: How to complete the missing resources?}
REST APIs operate on persistent resources, some created long ago.
Since servers typically retain logs only for a limited period (e.g., days or weeks), earlier resource creation records may be lost.
Without completing these resources, operation sequences---even those preserving business constraints---can fail to execute.

\item \textbf{Challenge-3: How to broaden the coverage of derived operation sequences?}
The operation sequences derived from HRLogs provide only limited coverage of the service.
\techname{} therefore uses these sequences as seeds to initiate a REST API fuzzing.
However, generating mutations that both preserve business constraints and explore diverse variations remains tricky.

\end{itemize}

To address these challenges, \techname{} incorporates three main designs: log slice generation (\textbf{Challenge-1}), log slice enhancement (\textbf{Challenge-2}), and business-aware REST API fuzzing (\textbf{Challenge-3}).
First, to preserve clean and complete constraints, \techname{} applies a locality slicing strategy to partition HRLogs into smaller slices.
This strategy is motivated by the observation that operations belonging to the same functionality tend to act on overlapping resources and are often executed consecutively within a short interval.
At the same time, \techname{} extracts valid parameter combinations and values from HRLogs for constructing concrete requests in subsequent stages.
Next, \techname{} enhances the obtained log slices by adding slices for operations missing from HRLogs and completing missing but required resources within the slices using parameter-to-resource dependencies.
Finally, the enhanced slices are used as initial seeds for business-aware REST API fuzzing.
During fuzzing, \techname{} employs business-aware and fault-triggering mutators to generate new operation sequences that both comply with the extracted business constraints and maximize the likelihood of uncovering faults in REST APIs.

To thoroughly evaluate \techname{}, we compared it with eight REST API testing techniques (including state-of-the-art ones such as \arat{} \cite{kim2022automated}, \morest{} \cite{liu2022morest}, \evomaster{} \cite{arcuri2019evomaster}, and \deeprest{} \cite{corradini2024deeprest}) on 17 REST services (S01-S17).
Services S01-S10 are collected from the recently published RESTgym \cite{corradini2025restgym} benchmark.
Most functionalities in these services are business-insensitive, with only a small portion being business-sensitive.
In contrast, services S11-S17 come from the GitLab REST service, which contains a rich set of business-sensitive functionalities.
Specifically, S11-S16 are sub-services of the entire GitLab REST service, previously evaluated in studies \cite{atlidakis2019restler} and \cite{wu2022restct}; S17 is the entire GitLab REST service.
To the best of our knowledge, we are the first to evaluate existing REST API testing techniques on a complete GitLab REST service with over 1,000 API operations (prior evaluations only consider services with fewer than 100 operations).

Experimental results demonstrate that \techname{} consistently outperforms the other techniques in terms of operation coverage, line coverage, and bug detection.
On S01-S10, \techname{} achieves the highest operation coverage on nine services and improves line coverage by an average of 15.2\% over the best-performing technique, \arat{}.
On S11–S16, \techname{} improves operation coverage by 263.1\% and line coverage by 26.5\% compared with the best-performing technique, \restct{}.
On S17, \techname{} achieves 188.6\% higher operation coverage and 56.6\% higher line coverage than the best performing technique, \evomaster{}.
For bug detection, across all 17 services, \techname{} detects the most \texttt{5XX} bugs---108 in total (38 of them cannot be detected by other techniques).

In summary, this paper makes the following contributions:

\begin{itemize}[leftmargin=10pt]

\item \textbf{Innovative Technique.}
We propose \techname{}, a novel REST API testing technique that leverages historical request logs to effectively test the business-sensitive functionalities behind REST APIs.

\item \textbf{Extensive Evaluation.}
We evaluate \techname{} against eight baselines on 17 REST services with 4,840 CPU hours.
To the best of our knowledge, this is the first work to evaluate existing techniques on a large-scale service, which has more than 1,000 API operations.

\item \textbf{Practical Tool.}
We prototype \techname{} and release all code to support future research.

\end{itemize}

\section{Background} \label{sec:preliminaries}

\setlength{\textfloatsep}{10pt}
\begin{figure}
  \centering
  \setlength{\fboxsep}{0pt} 
  \setlength{\fboxrule}{0.2pt}
  
  \begin{minipage}[b]{0.4\textwidth}
    \centering
    \fbox{\includegraphics[width=\textwidth - 2\fboxrule]{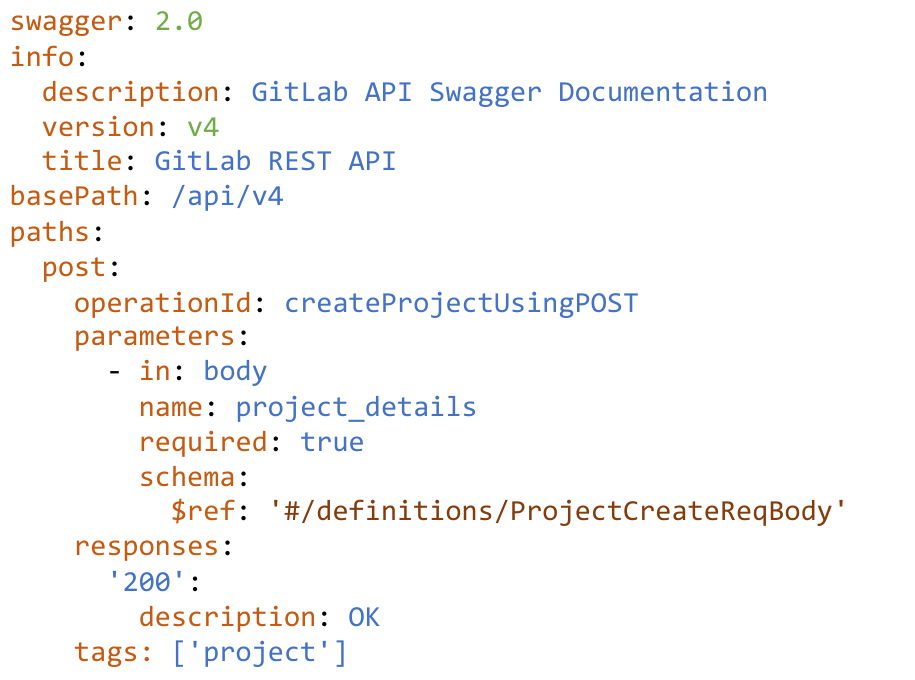}}
    \caption{A swagger specification snippet.}
    \label{fig:swagger}
  \end{minipage}
  \hspace{3em}
  \begin{minipage}[b]{0.45\textwidth}
    \centering
    \begin{subfigure}{\textwidth}
      \centering
      \fbox{\includegraphics[width=\textwidth - 2\fboxrule]{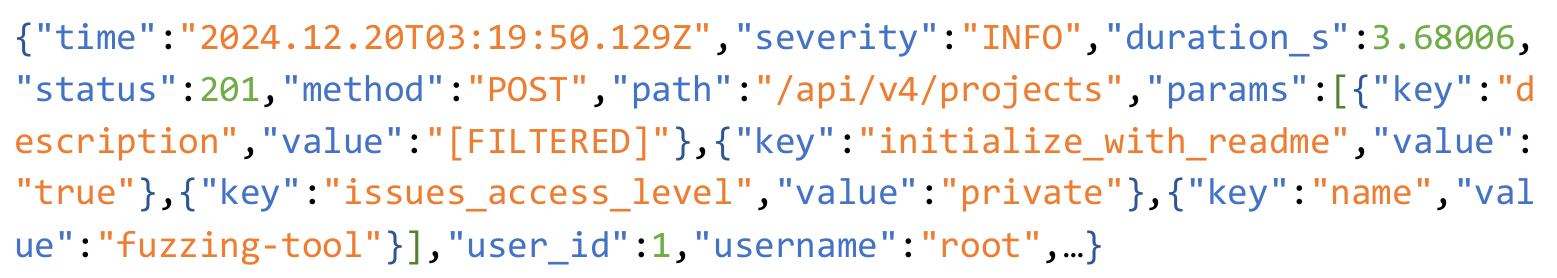}}
      \caption{An entry from Gitlab logs.}
      \label{fig:gitlab-log}
    \end{subfigure}
    
    \vspace{1em} 
    
    \begin{subfigure}{\textwidth}
      \centering
      \fbox{\includegraphics[width=\textwidth - 2\fboxrule]{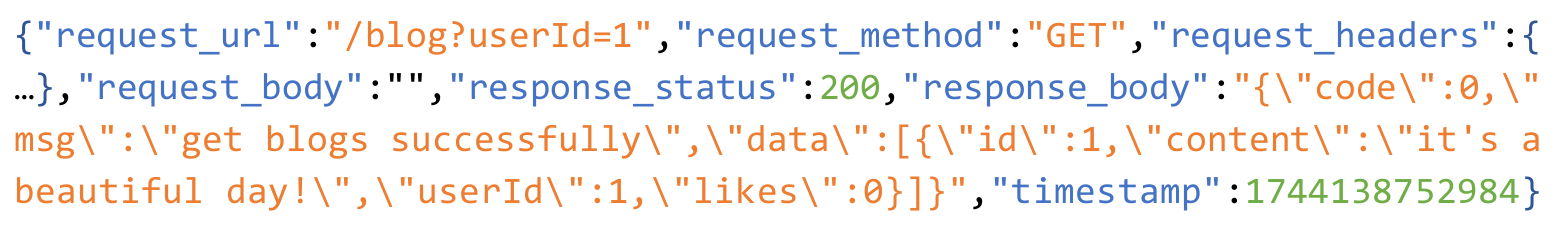}}
      \caption{An entry from Nginx logs.}
      \label{fig:nginx-log}
    \end{subfigure}
    
    \caption{An example of REST API HRLog entries.}
    \label{fig:right_combined}
    \Description{An example of REST API HRLog entries.}
  \end{minipage}
  \Description{}
\end{figure}
\noindent \textbf{REST APIs and Specifications.}
REST is a popular web API design style built on top of HTTP \cite{fielding2000architectural, berners1996hypertext}.
Services providing REST APIs are referred to as REST services.
An API in a REST service is also called an API operation.
These operations are used to manage persistent resources in services \cite{saleem2016quality}.
Each operation comprises a URI, an HTTP method (e.g., GET, POST, PUT, DELETE), and optional parameters.
Executing an operation means sending the corresponding HTTP request to the service.
Mainstream REST API specifications, such as OpenAPI \cite{openapi} and Swagger \cite{swagger}, describe how resources can be accessed and managed.
\autoref{fig:swagger} presents a Swagger specification snippet, where the header records service metadata and the \texttt{paths} field enumerates all available API operations.

\noindent \textbf{HRLogs.}
Historical request logs (HRLogs) record concrete invocations of API operations and are generated either by the service itself or by proxy gateways (e.g., Nginx\cite{nginx}).
In this paper, we collectively refer to logs from both sources as HRLogs.
HRLog entries record concrete historical requests, containing the timestamps, URIs, parameters, and response status codes \cite{he2021survey}.
\autoref{fig:gitlab-log} shows a log entry from the GitLab service itself, while \autoref{fig:nginx-log} presents one from the Nginx gateway.
\techname{} leverages HRLogs for REST API testing because these logs contain authentic historical executions of functionalities, including the order of operations and the parameter combinations/values used. 
This information inherently captures business constraints, which are missing from REST specifications, enabling \techname{} to preserve these constraints into operation sequences.

\noindent \textbf{REST API Testing.}
The primary goal of REST API testing is to trigger \texttt{5XX} response codes, which indicate server-side errors.
A test case consists of a sequence of API operations, such as \texttt{POST /projects} $\rightarrow$ \texttt{DELETE /projects/:id}.
In this sequence, the \texttt{id} used in the deletion operation is obtained from the response of the creation operation, representing a \emph{resource constraint}.
Existing studies \cite{atlidakis2019restler, wu2022restct, viglianisi2020resttestgen, arcuri2019evomaster, liu2022morest} focus on extracting such resource constraints from REST specifications to generate operation sequences, but often overlook critical \emph{business constraints}, resulting in insufficient testing of REST APIs.
To address this, \techname{} innovatively leverages HRLogs to generate operation sequences that preserve business constraints.
It further performs mutation-based fuzzing \cite{manes2019art, miller1990empirical, qian2024dipri, zhu2022fuzzing, qian2025funfuzz, bohme2016coverage, bohme2017directed} on these sequences to more thoroughly explore the REST service.

\section{Motivating Example} \label{subsec:motivating-example}
\begin{figure}[t]
    \centering
    \includegraphics[width=\linewidth]{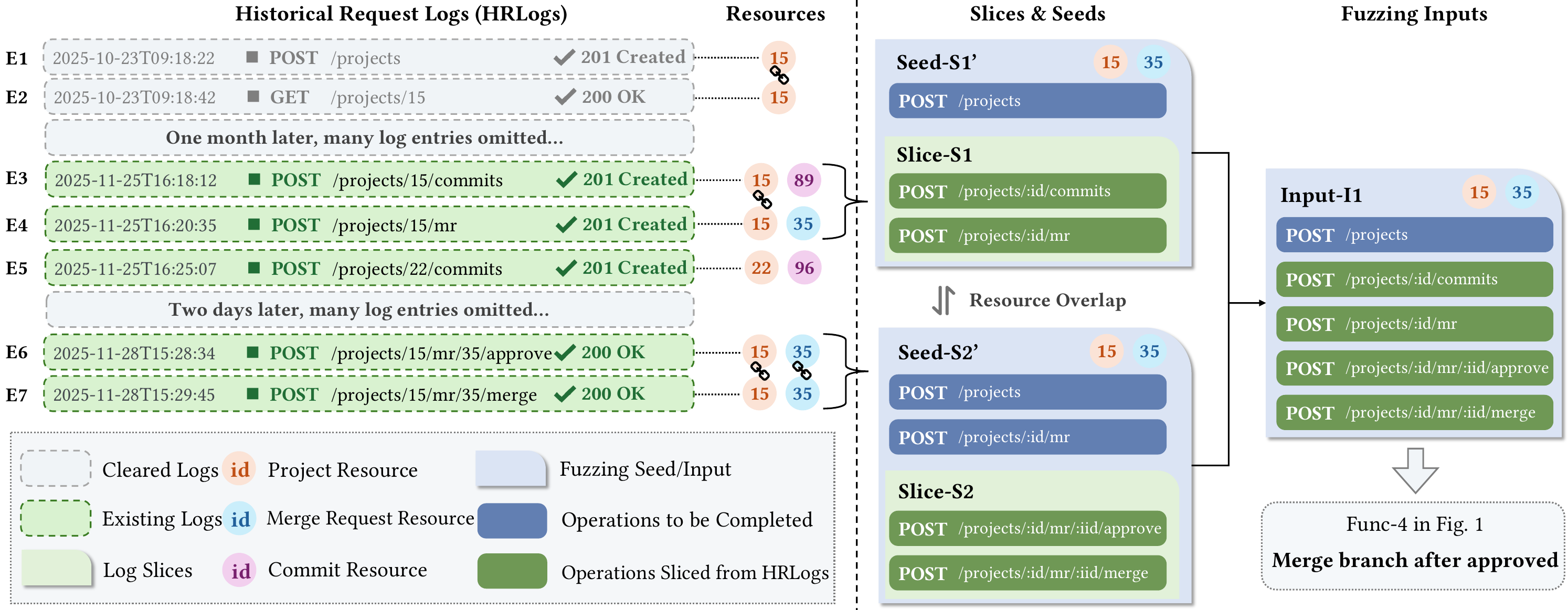}
    \caption{An example of \techname{} generating operation sequences to exercise Func-4 in \autoref{fig:business-functionality-example}.
    The left shows HRLogs and involved resources, while the right shows the generated log slices, seeds, and fuzzing inputs.}
    \label{fig:motivating-example}
    \Description{}
\end{figure}
This section illustrates how \techname{} constructs operation sequences to exercise the business-sensitive functionality Func-4 in \autoref{fig:business-functionality-example}.
To correctly exercise Func-4, operations must be executed in a strict order from OP-0 to OP-4:

\begin{tabular}{@{}c l l@{}} 
    $\bullet$ & OP-0 (create a project):               & \texttt{POST /projects} \\
    $\bullet$ & OP-1 (create a commit):                & \texttt{POST /projects/:id/commits} \\
    $\bullet$ & OP-2 (create a merge request):         & \texttt{POST /projects/:id/merge\_requests} \\
    $\bullet$ & OP-3 (approve the merge):              & \texttt{POST /projects/:id/merge\_requests/:iid/approve} \\
    $\bullet$ & OP-4 (merge the branch):               & \texttt{PUT \ /projects/:id/merge\_requests/:iid/merge} \\
\end{tabular}

\noindent
The absence of any operation or an incorrect execution order will cause the functionality to fail.

\textbf{Limitations of Specification-based Techniques.}
Existing REST API testing techniques \cite{atlidakis2019restler, hatfield2022schemathesis, karlsson2020quickrest, arcuri2019evomaster, viglianisi2020resttestgen, kim2022automated, liu2022morest, wu2022restct, corradini2024deeprest, kim2025llamaresttest} rely on REST specifications to generate operation sequences that satisfy resource constraints.
For example, they can produce sequences such as (OP-0, OP-1) and (OP-0, OP-2, OP-4), since OP-1,2 depend on OP-0, and OP-4 depends on OP-2.
However, they struggle to generate a complete sequence for Func-4 because the required business constraints are absent from the specifications.
In Func-4, the key business constraint is that both OP-3 and OP-4 must be present, and OP-3 (approve) must be executed before OP-4 (merge).
Without the business constraints, these techniques fail to adequately test business-sensitive functionalities behind REST APIs.

\textbf{Our Solution.}
\techname{} addresses this limitation by leveraging HRLogs.
\autoref{fig:motivating-example} depicts how \techname{} tackle with Func-4 using HRLogs.
The left side shows an HRLog piece excerpt from a GitLab service and the resources involved in each log entry.
Entries E3 and E4 create a commit (OP-1) and a merge request (OP-2) under project 15, while E6 and E7 approve (OP-3) and merge (OP-4) the request.
Although OP-1,2,3,4 occur in the exact order required by Func-4, which indicates that the business constraint of Func-4 is indeed captured in the HRLogs, the corresponding entries are scattered across chaotic logs, and the project-creation operation (OP-0) is missing due to log rotation (i.e., the mechanism that periodically clears old logs).
Therefore, to reconstruct a complete and valid operation sequence for Func-4, \techname{} performs three main steps.

\ding{182}
To preserve business constraints, \techname{} applies a locality-slicing strategy to partition HRLogs into smaller log slices.
Based on resource overlap and temporal proximity, entries E3 and E5 form slice S1, while E6 and E7 form slice S2.
\ding{183}
To ensure that each slice has proper resource constraints, \techname{} completes missing resource-creation operations using parameter-to-resource dependencies.
In S1, both operations require an existing project resource, inferred from their parameter \texttt{id} pointing to the same project (\texttt{id=15}).
Therefore, \techname{} prepends a project-creation operation to S1.
Similarly, it prepends a project-creation operation followed by a merge-request-creation operation to S2.
After completion, slices S1 and S2 become seeds S1' and S2'.
\ding{184}
To handle cases where operations of the same functionality are scattered across different seeds, \techname{} splices similar seeds during fuzzing.
Since S1' and S2' involve the same project and merge request, they are combined in chronological order.
To avoid redundant resource creation, splicing is applied to the original slices S1 and S2.
After splicing, resource completion is applied again to produce the final input I1, which preserves all the operations and a correct execution order (from OP-0 to OP-4) of Func-4.

By constructing such operation sequences that preserve business constraints, \techname{} enables deeper exploration of business-sensitive functionalities and further mutates these sequences to uncover unexpected behaviors and bugs in REST APIs.

\section{Approach} \label{sec:approach}
\setlength{\textfloatsep}{10pt}
\begin{figure}
    \centering
    \includegraphics[width=\linewidth]{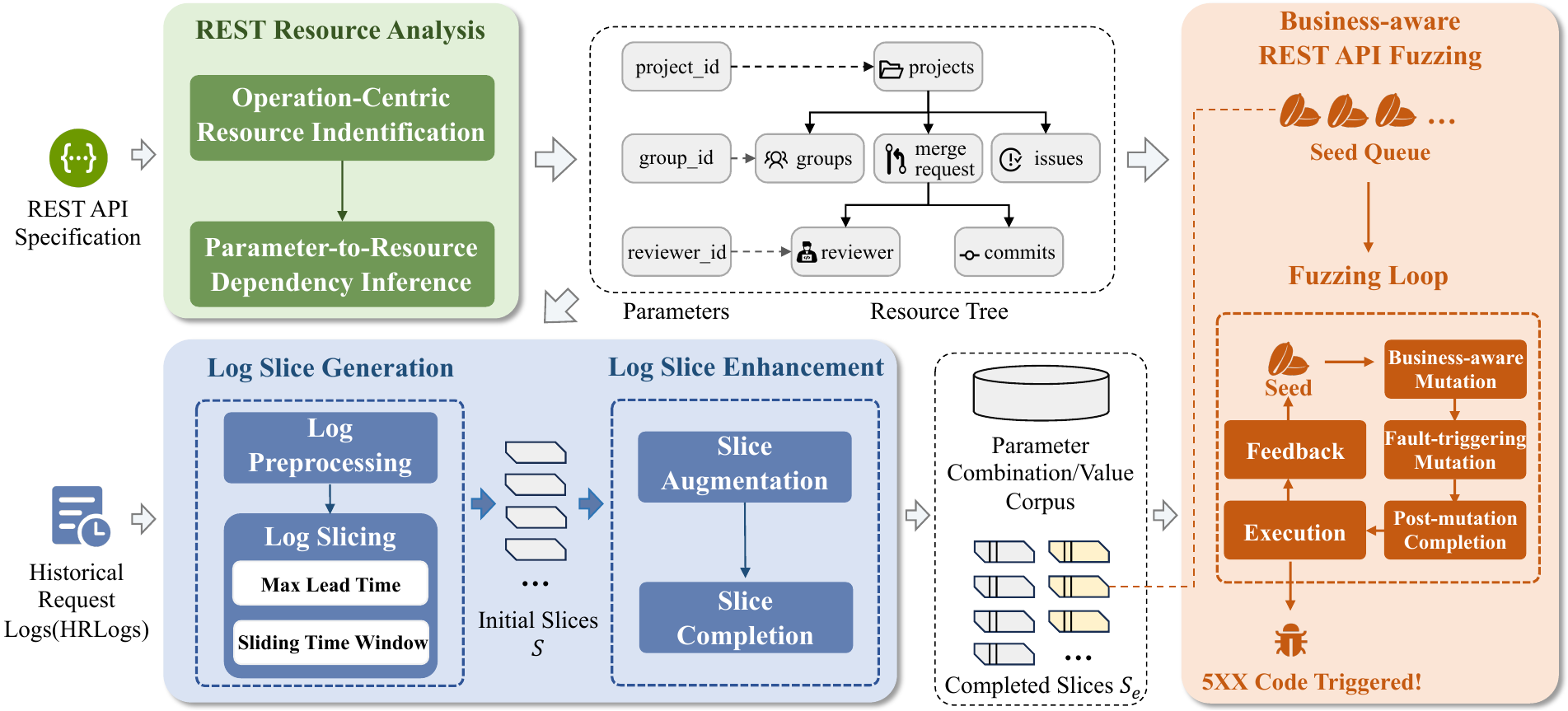}
    \caption{The overview of \techname{}}
    \label{fig:overview}
    \Description{}
\end{figure}
\autoref{fig:overview} presents an overview of \techname{}, which comprises four stages: \emph{REST resource analysis}, \emph{log slice generation}, \emph{log slice enhancement}, and \emph{business-aware REST API fuzzing}.

\noindent
\textbf{REST Resource Analysis.}
The first stage focuses on extracting REST resource information required by the subsequent three stages.
Specifically, \techname{} leverages a Large Language Model (LLM) to identify the \emph{resources} managed by the service and then infers the \emph{parameter-to-resource dependencies}.
Resources act as key cues for grouping operations that belong to the same business-sensitive functionality, while parameter-to-resource dependencies ensure that each generated operation sequence satisfies the required resource constraints.
The details are in \autoref{subsec:resource-analysis}.

\noindent
\textbf{Log Slice Generation.}
Directly using raw HRLogs for testing would intermingle unrelated operations and functionalities, making testing unscalable and uncontrollable.
Therefore, \techname{} partitions HRLogs into shorter log slices in this stage.
To preserve business constraints within each slice, \techname{} adopts a locality-slicing strategy.
This strategy ensures that log entries within the same slice operate on overlapping resources and are temporally close, making them more likely to belong to the same business-sensitive functionality.
Further details are described in \autoref{subsec:log-slice-gen}.

\noindent
\textbf{Log Slice Enhancement.}
Since HRLogs only record operations used by users in history and older records are periodically deleted, the previously obtained slices may have two drawbacks: (1) some operations under test may be missing from the slices, and (2) resource-creation operations in each slice may be absent.
\techname{} addresses this in two steps.
First, it creates a new slice for each missing operation.
Next, it leverages parameter-to-resource dependencies to complete any missing resource-creation operations in all slices, ensuring each slice forms a valid and effective operation sequence.
Further details about this stage are described in \autoref{subsec:log-slice-enhance}.

\noindent
\textbf{Business-aware REST API Fuzzing.}
The number of slices generated from HRLogs is limited.
To broaden their testing impact, \techname{} uses the slices as initial seeds and performs business-aware fuzzing.
It employs two types of mutators: \emph{business-aware mutators} and \emph{fault-triggering mutators}.
Business-aware mutators generate new operation sequences while preserving business constraints, enabling deeper exploration of the target service.
Fault-triggering mutators apply more aggressive mutations on the valid sequences produced by business-aware mutators, aiming to expose unexpected bugs.
The details are presented in \autoref{subsec:baraf}.

\subsection{REST Resource Analysis} \label{subsec:resource-analysis}
As the initial stage, REST resource analysis prepares the essential information required by the subsequent stages: \emph{resources} and \emph{parameter-to-resource dependencies}.
Resources are used to identify operations belonging to the same business-sensitive functionality, while parameter-to-resource dependencies support the completion of missing resources.
To derive them, \techname{} takes two steps: operation-centric resource identification and parameter-to-resource inference.

\subsubsection{Operation-centric Resource Identification} \label{subsubsec:operation-centric-resource-identification}
\begin{figure}
    \centering
    \begin{subfigure}[b]{0.32\textwidth}
        \centering
        \includegraphics[width=\linewidth]{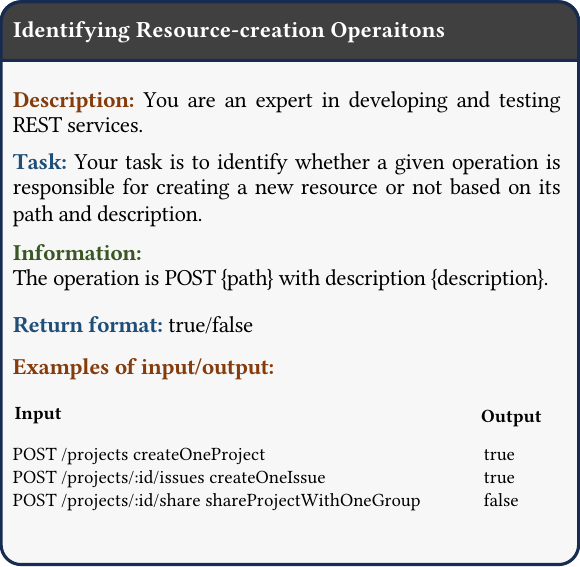}
        \caption{Prompt template for identifying resource-creation operations.}
        \label{fig:prompt-post}
    \end{subfigure}
    \hfill 
    \begin{subfigure}[b]{0.32\textwidth}
        \centering
        \includegraphics[width=\linewidth]{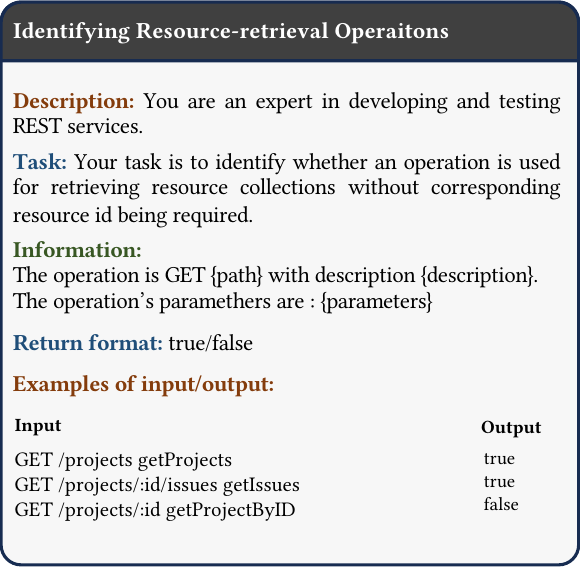}
        \caption{Prompt template for identifying resource-retrieval operations.}
        \label{fig:prompt-get}
    \end{subfigure}
    \hfill
    \begin{subfigure}[b]{0.32\textwidth}
        \centering
        \includegraphics[width=\linewidth]{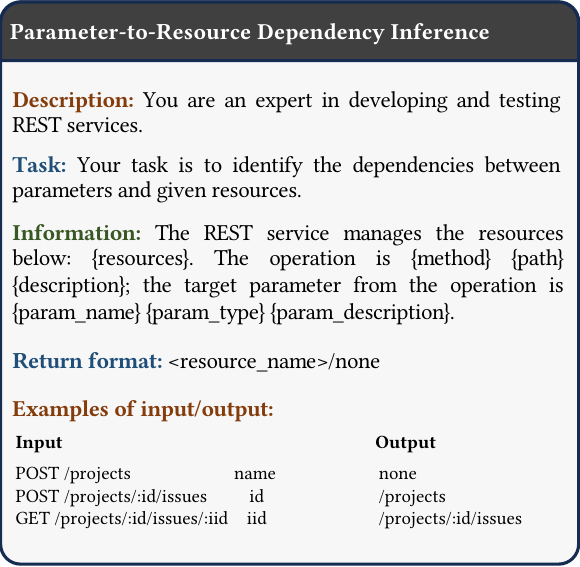}
        \caption{Prompt template for parameter-to-resource dependency inference.}
        \label{fig:prompt-dependency}
    \end{subfigure}

    \caption{Prompt templates for REST resource analysis.}
    \label{fig:prompts}
    \Description{Description for accessibility}
\end{figure}
In addition to constructing valid sequences, resources also help identify business constraints: operations belonging to the same business-sensitive functionality typically operate on overlapping resources.
However, resources are not explicitly specified in REST API specifications, which primarily describe operations.
Therefore, to obtain resources managed by the service, \techname{} adopts an \emph{operation-centric} approach.

Specifically, \techname{} infers resources by analysing the semantics of API operations.
In practice, not all \texttt{POST} operations create resources, nor do all \texttt{GET} operations retrieve resources.
To address this, \techname{} iterates over the \texttt{POST} and \texttt{GET} operations and uses an LLM to determine which operations create or retrieve resources (\autoref{fig:prompt-post}, \autoref{fig:prompt-get}). 
For \texttt{POST} operations, it selects those whose semantics indicate the creation of new resources (e.g., \texttt{POST /projects}) while excluding operations that merely act on existing resources (e.g., \texttt{POST /projects/:id/share}).
For \texttt{GET} operations, it focuses on those that retrieve collections of resources without requiring resource identifiers (e.g., \texttt{GET /projects}), excluding operations that access individual resources (e.g., \texttt{GET /projects/:id}).
Finally, \techname{} constructs a resource set $\mathcal{R}$, deriving each resource's name from the operation path.
These resources are further organized into a hierarchical resource tree: a resource whose name is a complete prefix of another is treated as the parent.
For example, resource \texttt{/projects} is identified as the parent of resource \texttt{/projects/:id/issues}.

\subsubsection{Parameter-to-resource Dependency Inference} \label{subsubsec:dependency-infer}
Parameter-to-resource dependencies are a critical basis for generating valid operation sequences.
They are used in subsequent log slice completion (\autoref{subsubsec:slice-completion}) and fuzzing (\autoref{subsec:baraf}) stages.
To extract these dependencies, \techname{} iterates over all parameters and leverages an LLM to determine whether a parameter originates from any resource in the previously obtained resource set $\mathcal{R}$.
The LLM is prompted with the operation context, target parameter, and list of known resources, returning either the dependent resource name or None (\autoref{fig:prompt-dependency}).
For example, parameter \texttt{name} in \texttt{POST /projects} does not depend on any existing resource because it is used to create a new project; in contrast, parameter \texttt{id} in \texttt{POST /projects/:id/issues} depends on the \texttt{/projects} resource since it references an existing project under which a new issue can be created.
By systematically applying this approach, \techname{} constructs a comprehensive mapping of parameters to resources.

\subsection{Log Slice Generation} \label{subsec:log-slice-gen}

Real-world HRLogs are typically chaotic: they contain requests from multiple users, span long time periods, and cover heterogeneous functionalities.
As a result, directly using these logs for testing (e.g., replaying entire logs) is problematic, as it mixes unrelated operations and functionalities, leading to uncontrollable and unscalable testing.
To address this, \techname{} partitions HRLogs into smaller slices, each serving as an independent operation sequence for testing.
Specifically, \techname{} first preprocesses these HRLogs by removing irrelevant requests and constructing user-independent queues; then, it employs the locality-slicing strategies to generate log slices that preserve business constraints.

\subsubsection{Log Preprocessing} \label{subsubsec:log-preprocessing}
In this phase, \techname{} takes four steps to prepare for log slicing.
First, \techname{} removes any operations not declared in the specification, as such operations are typically from other services and are less relevant to the service under test.
It also filters out invalid requests with \texttt{4XX} response codes.
Then, \techname{} extracts parameter combinations and values from requests with \texttt{2XX} response codes, storing them into a corpus for subsequent request construction.
Next, \techname{} identifies the resource instances involved in each request based on the parameter-to-resource dependencies obtained in \autoref{subsubsec:dependency-infer}.
A resource instance refers to a concrete occurrence of a resource observed in the HRLogs, identified by a specific value (e.g., an ID) extracted from requests.
Finally, to avoid interference among different users, \techname{} splits HRLogs into user-independent queues for each user based on user identifiers (e.g., user IDs, tokens, or cookies).


Each user-independent queue is an ordered collection of \emph{log entries}, where each log entry corresponds to a request record in the HRLogs.
Formally, a log entry $e$ can be formalized as a quintuple $e = \langle t_e, o_e, \mathcal{P}_e ,\mathcal{I}_e, \Phi_e \rangle$ where $t_e$ is the entry timestamp, $o_e$ is the executed operation, $\mathcal{P}_e$ is the used parameters, $\mathcal{I}_e$ is the resource instances involved in the entry, and \(\Phi_e : \mathcal{P}_e \to \mathcal{I}_e \cup \{\text{None}\}\) is a mapping from each parameter $p$ of $o_e$ to the corresponding resource instance in $\mathcal{I}_e$.

\subsubsection{Log Slicing} \label{subsubsec:log-slicing}
In this phase, \techname{} partitions each user-independent queue into smaller log slices.
Each log slice contains a sequence of log entries and serves as an operation sequence for testing.
Regarding preserving business constraints during slicing, our key observation is that operations belonging to the same business-sensitive functionality tend to operate on overlapping resource instances, and some of these operations are typically executed consecutively within a short interval.
Therefore, a log slice that preserves business constraints should satisfy two properties: (1) its entries involve overlapping resource instances, and (2) its entries are temporally close.
To generate log slices that meet these properties, \techname{} adopts two complementary locality-slicing strategies: \emph{maximum lead time slicing (MLTS)} and \emph{sliding time window slicing (STWS)}.



MLTS emphasizes minimizing the temporal gaps between consecutive entries in a slice.
Formally, for a candidate slice $\sigma = \{e_1, \dots, e_m\}$ ordered by timestamp $t_{e_1} < \dots < t_{e_m}$, we require 
\[
\max_{i=2}^{m} (t_{e_i} - t_{e_{i-1}}) \le \Delta t_\mathit{mlt},
\]
where $\Delta t_\mathit{mlt}$ is the maximum allowable lead time between consecutive entries.
This ensures that all entries in the slice occur in rapid succession, capturing tightly coupled operations.

STWS focuses on bounding the overall time span of a slice. 
Given a window size $\Delta t_\mathit{stw}$, slices $\sigma$ are formed such that
\[
t_\text{max}(\sigma) - t_\text{min}(\sigma) \le \Delta t_\mathit{stw},
\]
where $t_\text{min}(\sigma)$ and $t_\text{max}(\sigma)$ are the earliest and latest timestamps of entries in $\sigma$. 
This guarantees that the entire slice corresponds to a short interval of activity.

\begin{figure} 
\centering
\footnotesize 
\begin{minipage}[t]{0.49\textwidth} 
    \vspace{0pt} 
    \begin{algorithm}[H]
        \caption{\footnotesize Maximum Lead Time Slicing (MLTS)}
        \label{algo:mlts}
        \begin{algorithmic}[1]
            \Require Log entry queue $\mathcal{E}$, MLT threshold $\Delta t_\text{MLT}$
            \Ensure Set of log slices $S_k$
            \State Initialize $S_k, I, V \gets \emptyset$, $\sigma \gets []$, $Q \gets \{0\}$
            \While{$Q \neq \emptyset$}
                \State $s \gets Q.\text{pop()}, I \gets \mathcal{I}_{e_s}, t_\text{prev} \gets t_{e_s}, \sigma.\text{add}(e_s)$
                \For{$i = s+1$ \textbf{to} $|\mathcal{E}_k|$}
                    \State $t_\text{curr} \gets t_{e_i}$
                    \If{$t_\text{curr} - t_\text{prev} > \Delta t_\text{MLT}$}
                        \State $Q.\text{push}(i)~if~e_i \notin V$, \textbf{break}
                    \ElsIf{$\mathcal{I}_{e_i} \cap I = \emptyset$}
                        \State $Q.\text{push}(i)~if~e_i \notin V$
                    \Else
                        \State $\sigma.\text{add}(e_i), I \gets I \cup \mathcal{I}_{e_i}, t_\text{prev} \gets t_\text{curr}$
                    \EndIf
                \EndFor
                \State \textbf{if} $\sigma \neq \emptyset$ \textbf{then} $S_k \gets S_k \cup \{\sigma\}, V \gets V \cup\sigma$ 
                \State \textbf{end if}
            \EndWhile
            \State \Return $S_k$
        \end{algorithmic}
    \end{algorithm}
\end{minipage}
\hfill
\begin{minipage}[t]{0.49\textwidth} 
    \vspace{0pt}
    \begin{algorithm}[H]
        \caption{\footnotesize Sliding Time Window Slicing (STWS)}
        \label{algo:stws}
        \begin{algorithmic}[1]
            \Require Log entry queue $\mathcal{E}$, window size $\Delta t_\text{STW}$
            \Ensure Set of log slices $S_k$
            \State Initialize $S_k, I,V \gets \emptyset$, $\sigma \gets []$, $Q \gets \{0\}$
            \While{$Q \neq \emptyset$}
                \State $s \gets Q.\text{pop()}, I \gets \mathcal{I}_{e_s}, t_{\text{win}} \gets t_{e_s}, \sigma.\text{add}(e_s)$
                \For{$i = s+1$ \textbf{to} $|\mathcal{E}_k|$}
                    \If{$t_{e_i} - t_{\text{win}} > \Delta t_\text{STW}$}
                        \State $Q.\text{push}(i)~if~e_i \notin V$, \textbf{break}
                    \ElsIf{$\mathcal{I}_{e_i} \cap I = \emptyset$}
                        \State $Q.\text{push}(i)~if~e_i \notin V$
                    \Else
                        \State $\sigma.\text{add}(e_i)$, $I \gets I \cup \mathcal{I}_{e_i}$
                    \EndIf
                \EndFor
                \If{$\sigma \neq \emptyset$}
                    \State $S_k \gets S_k \cup \{\sigma\}, V \gets V \cup\sigma$
    \EndIf
\EndWhile
\State \Return $S_k$
\end{algorithmic}
    \end{algorithm}
\end{minipage}
\Description{}
\end{figure}
Algorithm~\ref{algo:mlts} and Algorithm~\ref {algo:stws} present the procedures of MLTS and STWS.
Both algorithms take a user-independent queue and a temporal threshold as input, and return a set of log slices $S_k$.
Initially, $S_k$, the current slice $\sigma$, the shared resource instance set $I$, and the set of already sliced entries $V$ are empty, and a queue $Q$ of potential slice starting indices is initialized (Line 1 in both).
For each index in $Q$, MLTS scans subsequent entries in temporal order.
If the interval between the current entry and the last entry in $\sigma$ exceeds the MLTS threshold, the current slice is finalized, and the current entry’s index is enqueued into $Q$ as a new starting point if it is not in $V$ (Lines~6--7,14--15 in Algorithm~\ref{algo:mlts}).
Otherwise, the algorithm checks for resource overlap: if the entry shares any resources with $\sigma$, it is appended, and its instances are merged into $I$; if not, the entry is skipped, but its index is enqueued for future slicing if not already processed (Lines~8–12 in Algorithm~\ref{algo:mlts}).
MLTS terminates when all indices in $Q$ have been processed.
STWS follows the same workflow, with the key difference that it measures the interval from the current entry to the slice’s starting entry, rather than between consecutive entries as in MLTS (Line 5 in Algorithm~\ref{algo:stws}).
After processing all the queues with the slicing algorithms, \techname{} merges the resulting slice sets to form the set of initial log slices $S$.

Note that when there are long time intervals between operations of a business-sensitive functionality, the locality-slicing strategy may split the corresponding log entries into different log slices.
Executing any single slice alone is therefore insufficient to exercise the full functionality.
To address this, \techname{} splices slices during the subsequent fuzzing stage (\autoref{subsec:baraf}) based on resource similarity, effectively reconstructing the complete operation sequence.

\subsection{Log Slice Enhancement} \label{subsec:log-slice-enhance}
HRLogs only record operations that have been executed by users, and older log entries are periodically cleared.
Therefore, the initial log slices $S$ derived from such logs suffer from two deficiencies: (1) they may not cover all service operations, and (2) some slices may lack required resource-creation operations.
To address these issues, \techname{} enhances $S$ in two steps. 
First, it adds new slices to $S$, each containing an operation that does not appear in the HRLogs. 
Second, it performs resource-consistency completion on each slice to ensure that all required resources are available and the slice can be executed successfully.

\subsubsection{Slice Augmentation}
To ensure that all operations of the service are represented in the log slices, \techname{} first identifies operations that do not appear in any of the initial slices.
For each missing operation, it constructs a corresponding log entry and adds it to the slice set as a single-entry slice.
This approach ensures that, in the subsequent resource-consistency completion stage, there is no need to distinguish between slices derived from HRLogs and those added to cover missing operations.
Finally, \techname{} obtains the augmented slice set $S_{aug}$.

\begin{algorithm}
\small
\caption{Resource-Consistency Slice Completion (RCSC)}
\label{algo:rcsc}
\begin{algorithmic}[1]
\Require A log slice $\sigma$
\Ensure Completed slice $\sigma'$ and parameter-to-entry mapping $\Phi_{\sigma'}$

\State Initialize $\mathcal{E}_{\text{new}} \gets []$ \Comment{list of newly created entries}
\State Initialize $\mathcal{I}_{\text{map}} \gets \{\}$ \Comment{mapping from resource instance $i \in \mathcal{I}$ to entry $e$}
\State Initialize $\Phi \gets \{\}$ \Comment{mapping from parameter $p$ to entry $e$}

\State $\mathcal{I}_\sigma \gets \bigcup_{e \in \sigma} \mathcal{I}_e$ \Comment{collect all resource instances in the slice}

\For{each $i \in \mathcal{I}_\sigma$ in descending resource-layer order} \Comment{\textcolor{blue}{Resource-creation entry construction}}
    \State $e_{\text{new}} \gets \textsc{createEntry}(i)$
    \State $\mathcal{I}_{\text{map}}[i] \gets e_{\text{new}}$, $\mathcal{E}_{\text{new}}.\text{adds}(e_{\text{new}})$
\EndFor

\State $\sigma' \gets \mathcal{E}_{\text{new}} \; \Vert \; \sigma$ \Comment{prepend new entries to the original slice}


\For{each entry $e \in \sigma'$} \Comment{\textcolor{blue}{Parameter-to-entry mapping construction}}
    \For{each parameter $p \in \mathcal{P}_e$ and $\Phi_e(p) \neq \text{None}$}
            \State $\Phi[p] \gets \mathcal{I}_{\text{map}}[\Phi_e(p)]$ \Comment{propagate dependencies}
    \EndFor
\EndFor

\State \Return $\sigma', \Phi_{\sigma'}$
\end{algorithmic}
\end{algorithm}
\subsubsection{Resource-consistency Slice Completion} \label{subsubsec:slice-completion}
REST services often enforce business constraints on resource-binding parameters, requiring two parameters to reference either the same or distinct resources.
For example, in GitLab, \texttt{POST /projects/:id/merge\_request} requires \texttt{source\_branch} and \texttt{target\_branch} to refer to different branches; associating both with the same branch would violate the functionality.
To satisfy such constraints, \techname{} performs \emph{Resource-Consistency Slice Completion} (RCSC), ensuring that each slice contains all required resource-creation operations while respecting resource usage.
Specifically, RCSC first prepends missing resource-creation operations to the slice following the resource hierarchy, then links parameters to their resource-creation operations using the instance IDs.

The procedure of RCSC is described in Algorithm~\ref{algo:rcsc}.
RCSC first collects all resource instances in the slice $\sigma$ (Line 4).
It then organizes them by their parent relationships and traverses them from higher to lower levels.
For each instance, \techname{} constructs a corresponding resource-creation entry, appends it to $\mathcal{E}_{\text{new}}$, and records the instance-to-entry mapping in $\mathcal{I}_{\text{map}}$ (Line 5--8).
After all required resource-creation entries are constructed, $\mathcal{E}_{\text{new}}$ is prepended to the original slice, yielding a completed slice $\sigma'$ (Line 9).
Finally, by composing the parameter-to-instance and instance-to-entry mappings, \techname{} derives a parameter-to-entry mapping $\Phi_{\sigma'}$(Line 10--14).


After processing all slices in the augmented slice set $S_{aug}$, \techname{} produces the completed slice set $S'$, in which each slice contains all required resource-creation operations.

\subsection{Business-aware REST API Fuzzing} \label{subsec:baraf}
Although the operation sequences in $S'$ preserve business constraints, they cover only a limited portion of the service behavior, resulting in insufficient exploration of the REST API.
To overcome this limitation, \techname{} performs a mutation-based fuzzing.
\techname{} treats each completed slice and its corresponding parameter-to-entry mapping as a fuzzing seed.
To explore more business functionalities while maximizing bug discovery, \techname{} employs two complementary categories of mutators: \emph{business-aware mutators} and \emph{fault-triggering mutators}.

Business-aware mutators aim to generate valid operation sequences that satisfy business constraints, intending to trigger feasible yet previously unobserved business functionalities.
A representative example is the \emph{Similar-Seed Splicing} mutator, which alleviates the fragmentation of business functionalities caused by locality-based slicing.
This mutator splices seeds that share overlapping resources according to their temporal order observed in HRLogs, thereby reassembling operations belonging to the same functionality while preserving execution-order constraints.
In addition, \techname{} includes mutators targeting constraints on parameter combinations and values.
These mutators replace parameter combinations or individual values using the corpus mined from HRLogs, producing inputs that more closely reflect real-world usage patterns and enabling exploration of diverse but valid business behaviors.
All sequences generated by business-aware mutators are re-completed before execution to ensure that newly introduced resource-binding parameters have their dependent resources available.
During execution, such parameters are dynamically assigned values based on the parameter-to-entry mapping in the seed.

In contrast, fault-triggering mutators apply more aggressive perturbations to the sequences produced by business-aware mutators.  
They randomly modify, add, or remove parameter values, insert or delete operations, and may even disable the runtime assignment of resource-binding parameters.
These disruptive mutations intentionally violate or stress business or resource constraints, aiming to expose latent faults and robustness issues in REST services.

\section{Evaluation} \label{sec:eval}

We evaluate \techname{} on the following research questions (RQs):

\begin{itemize}[leftmargin=12pt]
    \item \textbf{RQ1: Effectiveness on services with sparse businesses.}
    How does \techname{} compare with state-of-the-art REST API testing tools on lightweight services with sparse businesses?
    
    \item \textbf{RQ2: Effectiveness on services with dense businesses.}
    How does \techname{} compare with state-of-the-art REST API testing tools on industrial services with dense businesses?
    

    \item \textbf{RQ3: Contributions of core designs.}
    How do log slicing, log slice enhancement, and business-aware REST API fuzzing contribute to LoBREST’s testing performance?
\end{itemize}

\subsection{Experimental Setup} \label{subsec:exp-setup}

\subsubsection{Benchmarks.}
\begin{table}
\small
  \centering
  \caption{Selected Target REST Services.
  \# BMs refers to the number of business modules in a service.}

  \rowcolors{2}{gray!10}{white}
    \begin{tabularx}{\textwidth}{lXrrrcll}
    \toprule
    \textbf{ID} & \textbf{REST Service} & \textbf{\# Ops} & \textbf{LoC} & \textbf{\# BMs} & \textbf{Auth} & \textbf{Language} & \textbf{Version} \\
    \midrule
    S01    & Features Service & 18    & 456 & 1   & \ding{55}    & Java & commit 3f086ff \\
    S02    & Genome Nexus & 23    & 5242 & 6  & \ding{55}    & Java & v2.0.3 \\
    S03    & LanguageTool & 2     & 36583 & 1 & \ding{55}    & Java & v6.6 \\
    S04    & Market & 13    & 1583 & 5  & \ding{51}   & Java & commit bda6ca2 \\
    S05    & NCS   & 6     & 275 & 1 & \ding{55}    & Java & from WFD 4.0.0 \\
    S06    & SCS   & 11    & 295 & 1 & \ding{55}    & Java & from WFD 4.0.0 \\
    S07    & Project Tracking & 59    & 1298 & 7  & \ding{55}    & Java & commit b236c3a \\
    S08    & Person Controller & 12    & 211 & 1 & \ding{55}    & Java & commit 7b42660 \\
    S09    & REST Countries & 22    & 538 & 1 & \ding{55}    & Java & v2.0.5 \\
    S10   & User Management & 22    & 736 & 5 & \ding{55}    & Java & commit 04500e7 \\
    S11   & GitLab-Branch & 9  & \text{--} & \text{--} & \ding{51}   & Ruby & v18.4.5-ce \\
    S12   & GitLab-Commit & 15  & \text{--} & \text{--} & \ding{51}   & Ruby & v18.4.5-ce \\
    S13   & GitLab-Group & 17  & \text{--} & \text{--} & \ding{51}   & Ruby & v18.4.5-ce \\
    S14   & GitLab-Issue & 27  & \text{--} & \text{--} & \ding{51}   & Ruby & v18.4.5-ce \\
    S15   & GitLab-Project & 31  & \text{--} & \text{--} & \ding{51}   & Ruby & v18.4.5-ce \\
    S16   & GitLab-Repository & 10  & \text{--} & \text{--} & \ding{51}   & Ruby & v18.4.5-ce \\
    S17   & GitLab & 1099  & 50032 & 109 & \ding{51}   & Ruby & v18.4.5-ce \\
    \bottomrule
    \end{tabularx}
  \label{tab:selected-services}
\end{table}
We select a total of 17 open-source REST services, with their detailed information summarized in \autoref{tab:selected-services}.
Services S01-S10 come from the RESTgym benchmark \cite{corradini2025restgym}.
Services S11-S16 are sub-services of the GitLab REST service and have been widely used in prior evaluations \cite{wu2022restct,atlidakis2019restler} (their total lines of code cannot be computed because the partition is at the API rather than the file level).
S17 is the entire GitLab REST service with all 1,099 API operations.
To the best of our knowledge, we are the first to evaluate REST API testing tools on such a large-scale service with over 1,000 API operations---previous studies only consider services with fewer than 100 operations.

\subsubsection{Baselines.}
We choose eight representative and applicable REST API testing tools, which are:
\begin{itemize}[leftmargin=12pt]
    \item \restler{} \cite{atlidakis2019restler} is the first stateful REST API fuzzer that generates test sequences by inferring producer-consumer dependencies and analyzing dynamic feedback from responses.
    \item \evomaster{} \cite{arcuri2019evomaster, arcuri2021evomaster} is a search-based testing tool supporting both black-box and white-box modes.
    Since all the other tools in our evaluation operate in a black-box manner, \evomaster{} is also used in black-box mode for fairness (denoted as \evo{}).
    \item \rtg{} (abbreviated as \textsc{RTG}) \cite{viglianisi2020resttestgen} analyzes dependencies and shared attributes among APIs and constructs an operation dependency graph to guide test case generation.
    \item \morest{} \cite{liu2022morest} is also a graph-based REST API testing tool that incorporates data schemas as nodes within the graph to enhance its capability.
    \item \restct{} \cite{wu2022restct} is the first systematic and fully automatic approach that adopts combinatorial testing to REST API testing.
    \item \schemathesis{} (abbreviated as \textsc{Schma}) \cite{hatfield2022schemathesis} adopts a property-based testing approach to automatically derives structure-aware tests from OpenAPI schemas to uncover complex defects in REST APIs.
    \item \arat{} \cite{kim2023adaptive} is an adaptive REST API testing technique that uses reinforcement learning to prioritize operations and parameters during operation sequence generation.
    \item \deeprest{} \cite{corradini2024deeprest} leverages curiosity-driven deep reinforcement learning to uncover implicit business logic and hidden constraints in REST APIs
\end{itemize}

\subsubsection{Evaluation Metrics.}
Following prior work \cite{atlidakis2019restler,liu2022morest,kim2023adaptive,corradini2024deeprest,zhang2023open,kim2022automated}, we use operation coverage, line coverage, bug detection, and statistical effect size as our evaluation metrics.
\begin{itemize}[leftmargin=12pt]
    \item \emph{Operation coverage} measures the extent to which the testing tool explores the REST APIs.
    An operation is considered covered if it produces a \texttt{2XX} status code
    \item \emph{Line coverage} measures how thoroughly a testing tool exercises a service’s business functionalities at the line level.
    It has been widely adopted in prior studies \cite{liu2022morest, atlidakis2019restler}.
    We use JaCoCo and Ruby TracePoint to collect the line coverage of Java and Ruby Projects.
    \item \emph{Bug detection} is indicated by \texttt{5XX} status codes observed during testing.
    Each bug is identified by the corresponding service, operation, status code, and error message.
    \item \emph{Effect size ($\hat{A}_{12}$)} quantifies the magnitude of performance differences between two tools using the Vargha–Delaney $\hat{A}_{12}$ statistic based on the Mann–Whitney U test. 
    Our one-tailed alternative hypothesis is that \techname{} achieves higher values than the baseline tool being compared.
\end{itemize}


\subsubsection{HRLog Preparation.}
\techname{} relies on HRLogs to guide testing.
To generate HRLogs, we follow a controlled procedure.
We recruit a group of participants and first brief them on the typical business scenarios of each service.
They then interact with the services via REST APIs for fixed durations---24 hours for S01–S16 and 72 hours for S17, which has a substantially larger API set.

\subsubsection{Experimental Procedure and Environments.}
To enable parallel testing while maintaining isolation between different testings, we adopt a containerized setup.
Services and testing tools are deployed in separate containers.
During parallel testing, the overall CPU and memory utilization are kept below 60\%.
All experiments are conducted on a server running Ubuntu 20.04.6 LTS, equipped with an Intel(R) Xeon(R) Silver 4316 CPU @ 2.30GHz (80 logical cores) and 512 GB of RAM.

\subsection{RQ1: Effectiveness on services with sparse businesses} \label{subsec:rq1}
We categorize Services S01–S10 as services with sparse businesses, as most functionalities in them are business-insensitive.
Evaluating \techname{} on such services provides a baseline for its performance.
We set a time budget of one hour for each tool run, which is widely adopted and examined in prior works \cite{kim2023adaptive, liu2022morest, atlidakis2019restler, corradini2025restgym, kim2022automated}.
To alleviate the impact of randomness, each tool run is repeated 20 times.
\begin{table}
  \centering
  \caption{Comparison of coverage results on all target services.
  For each baseline, the value outside parentheses denotes the
average operation/line coverage, while the value inside is the $\hat{A}_{12}$ effect size. 
Entries marked with “--” indicate cases where a tool crashes or lacks authentication when testing a service.
}
  \label{tab:coverage-comparison}

  \begin{subtable}{\textwidth}
\centering
  \caption{Results of operation coverage.}
    \resizebox{\textwidth}{!}{%
    \rowcolors{2}{gray!10}{white}
    \begin{tabular}{lrrrrrrrrr}
    \toprule
          & \textbf{\techname{}} & \textbf{\restler{}} & \textbf{\evo{}} & \textbf{\textsc{RTG}} & \textbf{\morest{}} & \textbf{\restct{}} & \textbf{\textsc{Schma}} & \textbf{\arat{}} & \textbf{\deeprest{}} \\
    \midrule
    \textbf{S01} & \textbf{18.0} & 11.0\textsubscript{(1.00)} & 12.1\textsubscript{(1.00)} & 10.4\textsubscript{(1.00)} & 11.1\textsubscript{(1.00)} & 8.6\textsubscript{(1.00)} & 6.0\textsubscript{(1.00)} & 17.8\textsubscript{(0.60)} & 11.2\textsubscript{(1.00)} \\
    \textbf{S02} & \textbf{23.0} & 15.0\textsubscript{(1.00)} & 18.8\textsubscript{(1.00)} & 11.3\textsubscript{(1.00)} & 22.9\textsubscript{(0.53)} & --   & 21.0\textsubscript{(1.00)} & 22.1\textsubscript{(0.80)} & 10.3\textsubscript{(1.00)} \\
    \textbf{S03} & \textbf{2.0} & 1.0\textsubscript{(1.00)} & \textbf{2.0}\textsubscript{(0.50)} & 1.0\textsubscript{(1.00)} & \textbf{2.0}\textsubscript{(0.50)} & --   & 1.2\textsubscript{(0.88)} & \textbf{2.0}\textsubscript{(0.50)} & 1.0\textsubscript{(1.00)} \\
    \textbf{S04} & \textbf{13.0} & 9.0\textsubscript{(1.00)} & 10.0\textsubscript{(1.00)} & 9.0\textsubscript{(1.00)} & --   & 9.0\textsubscript{(1.00)} & 10.8\textsubscript{(1.00)} & --   & 9.0\textsubscript{(1.00)} \\
    \textbf{S05} & \textbf{6.0} & \textbf{6.0}\textsubscript{(0.50)} & \textbf{6.0}\textsubscript{(0.50)} & 5.0\textsubscript{(1.00)} & --   & \textbf{6.0}\textsubscript{(0.50)} & \textbf{6.0}\textsubscript{(0.50)} & \textbf{6.0}\textsubscript{(0.50)} & \textbf{6.0}\textsubscript{(0.53)} \\
    \textbf{S06} & \textbf{10.0} & \textbf{10.0}\textsubscript{(0.50)} & \textbf{10.0}\textsubscript{(0.50)} & \textbf{10.0}\textsubscript{(0.50)} & \textbf{10.0}\textsubscript{(0.50)} & \textbf{10.0}\textsubscript{(0.50)} & \textbf{10.0}\textsubscript{(0.50)} & \textbf{10.0}\textsubscript{(0.50)} & \textbf{10.0}\textsubscript{(0.50)} \\
    \textbf{S07} & \textbf{56.8} & 20.0\textsubscript{(1.00)} & 43.1\textsubscript{(1.00)} & 25.1\textsubscript{(1.00)} & 41.5\textsubscript{(1.00)} & --   & 23.4\textsubscript{(1.00)} & 39.3\textsubscript{(1.00)} & 22.7\textsubscript{(1.00)} \\
    \textbf{S08} & 8.0 & 2.0\textsubscript{(1.00)} & 5.0\textsubscript{(1.00)} & 3.6\textsubscript{(1.00)} & \textbf{9.0}\textsubscript{(0.00)} & --   & 5.8\textsubscript{(1.00)} & 7.0\textsubscript{(1.00)} & 2.8\textsubscript{(1.00)} \\
    \textbf{S09} & \textbf{12.0} & --   & 9.0\textsubscript{(1.00)} & \textbf{12.0}\textsubscript{(0.50)} & \textbf{12.0}\textsubscript{(0.50)} & 8.1\textsubscript{(1.00)} & 7.2\textsubscript{(1.00)} & \textbf{12.0}\textsubscript{(0.50)} & 10.3\textsubscript{(0.75)} \\
    \textbf{S10} & \textbf{21.0} & --   & 16.9\textsubscript{(1.00)} & 13.2\textsubscript{(1.00)} & 17.0\textsubscript{(1.00)} & --   & 16.9\textsubscript{(1.00)} & 16.9\textsubscript{(1.00)} & 13.9\textsubscript{(1.00)} \\
    \textbf{S11} & \textbf{8.0} & 1.0\textsubscript{(1.00)} & 5.0\textsubscript{(1.00)} & 5.5\textsubscript{(1.00)} & --   & 4.4\textsubscript{(1.00)} & 4.8\textsubscript{(1.00)} & --   & 5.3\textsubscript{(1.00)} \\
    \textbf{S12} & \textbf{12.0} & 1.0\textsubscript{(1.00)} & 3.0\textsubscript{(1.00)} & 3.0\textsubscript{(1.00)} & --   & 3.0\textsubscript{(1.00)} & 2.7\textsubscript{(1.00)} & --   & 2.8\textsubscript{(1.00)} \\
    \textbf{S13} & \textbf{11.0} & 1.0\textsubscript{(1.00)} & 10.0\textsubscript{(1.00)} & 1.0\textsubscript{(1.00)} & --   & 1.0\textsubscript{(1.00)} & 3.2\textsubscript{(1.00)} & --   & 1.0\textsubscript{(1.00)} \\
    \textbf{S14} & \textbf{22.6} & 3.0\textsubscript{(1.00)} & 7.1\textsubscript{(1.00)} & 15.7\textsubscript{(0.99)} & --   & 22.5\textsubscript{(0.55)} & 6.0\textsubscript{(1.00)} & --   & 21.9\textsubscript{(0.68)} \\
    \textbf{S15} & \textbf{26.7} & 1.0\textsubscript{(1.00)} & 13.9\textsubscript{(1.00)} & 22.7\textsubscript{(1.00)} & --   & 20.6\textsubscript{(1.00)} & 5.3\textsubscript{(1.00)} & --   & 24.9\textsubscript{(0.81)} \\
    \textbf{S16} & \textbf{8.0} & 1.0\textsubscript{(1.00)} & 3.0\textsubscript{(1.00)} & 3.0\textsubscript{(1.00)} & --   & 3.0\textsubscript{(1.00)} & 3.0\textsubscript{(1.00)} & --   & 2.9\textsubscript{(1.00)} \\
    \textbf{S17} & \textbf{352.7} & --   & 122.2\textsubscript{(1.00)} & 47.9\textsubscript{(1.00)} & --   & --   & 26.1\textsubscript{(1.00)} & --   & 43.3\textsubscript{(1.00)} \\
    \bottomrule
    \end{tabular}%
    }
  \label{tab:operation-coverage}%
  \end{subtable}

  \begin{subtable}{\textwidth}
  \centering
  \caption{Results of line coverage.}
  \resizebox{\textwidth}{!}{
  \rowcolors{2}{gray!10}{white}
    \begin{tabular}{lrrrrrrrrr}
    \toprule
          & \textbf{\techname{}} & \textbf{\restler{}} & \textbf{\evo{}} & \textbf{\textsc{RTG}} & \textbf{\morest{}} & \textbf{\restct{}} & \textbf{\textsc{Schma}} & \textbf{\arat{}} & \textbf{\deeprest{}} \\
    \midrule
    \textbf{S01} & \textbf{370.0}   & 220.0\textsubscript{(1.00)} & 266.9\textsubscript{(1.00)} & 227.8\textsubscript{(1.00)} & 231.5\textsubscript{(1.00)} & 200.7\textsubscript{(1.00)} & 178.0\textsubscript{(1.00)} & 364.4\textsubscript{(0.82)} & 230.1\textsubscript{(1.00)} \\
    \textbf{S02} & \textbf{2384.8} & 1534.0\textsubscript{(1.00)} & 1994.2\textsubscript{(1.00)} & 1545.2\textsubscript{(1.00)} & 1613.8\textsubscript{(1.00)} & --   & 1538.3\textsubscript{(1.00)} & 1882.0\textsubscript{(1.00)} & 1475.5\textsubscript{(1.00)} \\
    \textbf{S03} & \textbf{11193.7} & 1264.0\textsubscript{(1.00)} & 9069.2\textsubscript{(1.00)} & 1264.0\textsubscript{(1.00)} & 9646.5\textsubscript{(1.00)} & --   & 2866.7\textsubscript{(1.00)} & 10861.6\textsubscript{(0.96)} & 1264.0\textsubscript{(1.00)} \\
    \textbf{S04} & \textbf{791.5} & 591.0\textsubscript{(1.00)} & 633.4\textsubscript{(1.00)} & 570.0\textsubscript{(1.00)} & --   & 580.0\textsubscript{(1.00)} & 661.4\textsubscript{(1.00)} & --   & 570.0\textsubscript{(1.00)} \\
    \textbf{S05} & \textbf{265.5} & 203.0\textsubscript{(1.00)} & 177.7\textsubscript{(1.00)} & 167.0\textsubscript{(1.00)} & --   & 235.0\textsubscript{(1.00)} & 258.4\textsubscript{(1.00)} & 257.9\textsubscript{(0.96)} & 227.3\textsubscript{(1.00)} \\
    \textbf{S06} & \textbf{263.0}   & 181.0\textsubscript{(1.00)} & 197.2\textsubscript{(1.00)} & 194.7\textsubscript{(1.00)} & 185.8\textsubscript{(1.00)} & 180.8\textsubscript{(1.00)} & 194.0\textsubscript{(1.00)} & 200.5\textsubscript{(1.00)} & 191.0\textsubscript{(1.00)} \\
    \textbf{S07} & \textbf{563.6} & 406.0\textsubscript{(1.00)} & 495.9\textsubscript{(1.00)} & 398.6\textsubscript{(1.00)} & 527.4\textsubscript{(1.00)} & --   & 468.2\textsubscript{(1.00)} & 517.4\textsubscript{(1.00)} & 401.6\textsubscript{(1.00)} \\
    \textbf{S08} & \textbf{168.0}   & 57.0\textsubscript{(1.00)} & 156.0\textsubscript{(1.00)} & 46.8\textsubscript{(1.00)} & 156.4\textsubscript{(0.97)} & --   & 156.9\textsubscript{(1.00)} & 156.0\textsubscript{(1.00)} & 42.6\textsubscript{(1.00)} \\
    \textbf{S09} & 328.8 & --   & 322.1\textsubscript{(1.00)} & 327.4\textsubscript{(0.90)} & 329.0\textsubscript{(0.40)} & 319.1\textsubscript{(1.00)} & 316.4\textsubscript{(1.00)} & \textbf{329.1}\textsubscript{(0.38)} & 322.1\textsubscript{(0.72)} \\
    \textbf{S10} & \textbf{662.6} & --   & 461.9\textsubscript{(1.00)} & 352.2\textsubscript{(1.00)} & 491.3\textsubscript{(1.00)} & --   & 474.4\textsubscript{(1.00)} & 475.4\textsubscript{(1.00)} & 356.4\textsubscript{(1.00)} \\
    \textbf{S11} & \textbf{607.8} & 284.0\textsubscript{(1.00)} & 539.8\textsubscript{(1.00)} & 543.9\textsubscript{(1.00)} & --   & 544.6\textsubscript{(1.00)} & 525.3\textsubscript{(1.00)} & --   & 527.2\textsubscript{(1.00)} \\
    \textbf{S12} & \textbf{712.0}   & 285.0\textsubscript{(1.00)} & 579.0\textsubscript{(1.00)} & 554.1\textsubscript{(1.00)} & --   & 576.8\textsubscript{(1.00)} & 518.2\textsubscript{(1.00)} & --   & 506.3\textsubscript{(1.00)} \\
    \textbf{S13} & 436.6 & 254.0\textsubscript{(1.00)} & \textbf{460.9}\textsubscript{(0.00)} & 229.0\textsubscript{(1.00)} & --   & 229.0\textsubscript{(1.00)} & 322.8\textsubscript{(1.00)} & --   & 229.0\textsubscript{(1.00)} \\
    \textbf{S14} & \textbf{812.5} & 425.0\textsubscript{(1.00)} & 652.8\textsubscript{(1.00)} & 703.8\textsubscript{(1.00)} & --   & 739.1\textsubscript{(1.00)} & 620.7\textsubscript{(1.00)} & --   & 716.4\textsubscript{(1.00)} \\
    \textbf{S15} & \textbf{780.0}   & 354.0\textsubscript{(1.00)} & 585.4\textsubscript{(1.00)} & 668.4\textsubscript{(1.00)} & --   & 691.0\textsubscript{(1.00)} & 402.9\textsubscript{(1.00)} & --   & 686.9\textsubscript{(1.00)} \\
    \textbf{S16} & \textbf{592.6} & 282.0\textsubscript{(1.00)} & 526.5\textsubscript{(1.00)} & 520.4\textsubscript{(1.00)} & --   & 537.2\textsubscript{(1.00)} & 480.7\textsubscript{(1.00)} & --   & 481.4\textsubscript{(1.00)} \\
    \textbf{S17} & \textbf{3945.8} & --  & 2519.1\textsubscript{(1.00)} & 1630.5\textsubscript{(1.00)} & --   & -- & 1255.3\textsubscript{(1.00)} & --   & 1592.2\textsubscript{(1.00)}  \\
    \bottomrule
    \end{tabular}%
    }
  \label{tab:code-coverage}%
  \end{subtable}

\end{table}
\subsubsection{Operation Coverage} \label{subsubsec:rq1-1}
The operation coverage for Services S01-S10 is summarized in \autoref{tab:operation-coverage}.
\techname{} achieves the highest coverage on 9 out of the 10 services; on S08, it ranks second, trailing \morest{} by only one operation.
Most $\hat{A}_{12}$ values are at least 0.80, indicating consistent advantages across 20 runs.
$\hat{A}_{12}$ Values of 0.50 occur when \techname{} and baselines achieve identical coverage, typically on services with very simple functionality.
For example, Services S05 and S06 manage no persistent resources and expose only GET operations for basic numeric computation and string concatenation; consequently, each functionality involves only a single operation.
However, identical operation coverage does not imply equivalent effectiveness: as shown later in \autoref{subsubsec:rq1-2}, \techname{} still outperforms all baselines on S05 and S06 in terms of line coverage.

As for the average coverage rate, \techname{} consistently outperforms the eight baselines.
Specifically, \techname{} achieves an operation coverage of 90.4\%, reaching 100.0\% coverage on five services (S01-S05).
Compared with the top-performing baselines—\arat{} (82.6\%), \morest{} (78.6\%), and \evo{} (75.0\%)—\techname{} improves the average coverage by 9.4\%, 15.0\%, and 20.5\%.

\subsubsection{Line Coverage} \label{subsubsec:rq1-2}
The line coverage results for Services S01–S10 are summarized in \autoref{tab:code-coverage}.
\techname{} attains the highest line coverage on 9 of the 10 services; on the remaining one, its coverage is comparable to the best-performing baseline \arat{}, with a difference of only 0.3 lines.
The $\hat{A}_{12}$ values confirm that \techname{} consistently outperforms the baselines.
Overall, \techname{} attains an average line coverage of 66.6\%, improving by 15.2\% and 30.6\% over the strongest baselines, \arat{} and \morest{}.
A manual inspection of the source code reveals that, for several services (e.g., S01, S05-S06, S07, and S10), \techname{} already reaches the maximum line coverage, as the remaining uncovered code is either dead or unreachable via REST APIs.

\begin{figure}
    \centering
    \includegraphics[width=\linewidth]{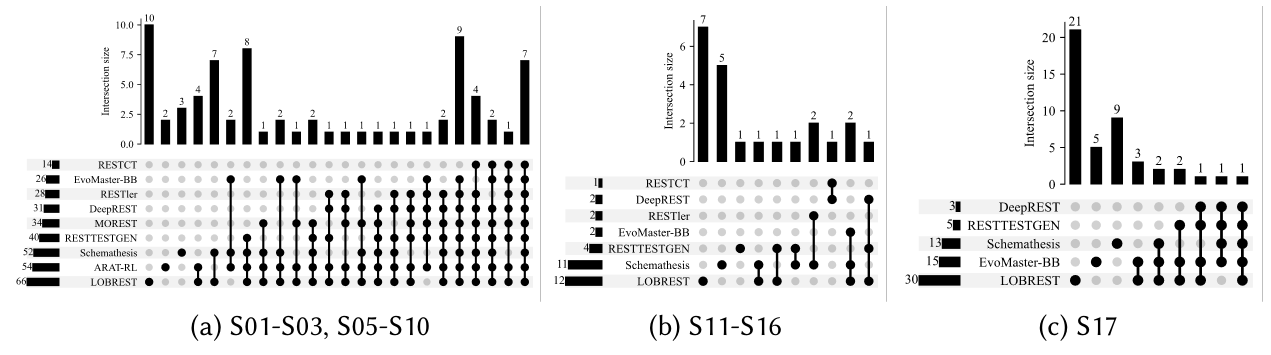}
    \captionsetup{skip=8pt}
    \caption{UpSet plots illustrating the bugs detected by all tools across Service S01-S17 (excluding S04 for fairness).
    Only bugs triggered at least 10 times across 20 runs are counted.}
    \label{fig:bugs}
    \Description{}
\end{figure}

\subsubsection{Bug Detection} \label{subsubsec:rq1-3}
\autoref{fig:bugs}a shows the overlaps of bugs found by \techname{} and eight baseline tools.
For fairness, Service S04 is excluded because the SOTA tools \arat{} and \morest{} could not run due to authentication issues.
The results show that \techname{} detects the most bugs (66), followed by \arat{} (54), \schemathesis{} (52), and \rtg{} (40); the remaining baselines detect fewer than 40.
Moreover, the matrix panel at the bottom of the UpSet plot reveals that only \techname{}, \arat{}, and \schemathesis{} can find bugs that other tools fail to detect, with \techname{} performing the best (10 vs. 3 and 2).
Totally, 74 bugs are found across these nine services, with \techname{} responsible for 89.2\%.
In summary, \techname{} consistently uncovers most bugs in Services S01–S10 (excluding S04) and identifies more independent bugs than other tools.

\begin{summarybox}
    \textbf{Answers to RQ1:} LoBREST demonstrates strong baseline performance on services with sparse businesses.
    It achieves the highest operation coverage on 9 out of the 10 services, reaching 100.0\% operation coverage on 5 services.
    LoBREST improves average line coverage by 15.2\% compared with the second-best tool \arat{}.
    It also detects the most bugs---66 in total---22.0\% more than the second-best tool \arat{}.
\end{summarybox}

\begin{figure}
    \centering
    \begin{minipage}[b]{0.46\textwidth}
        \centering
        \includegraphics[height=4.1cm]{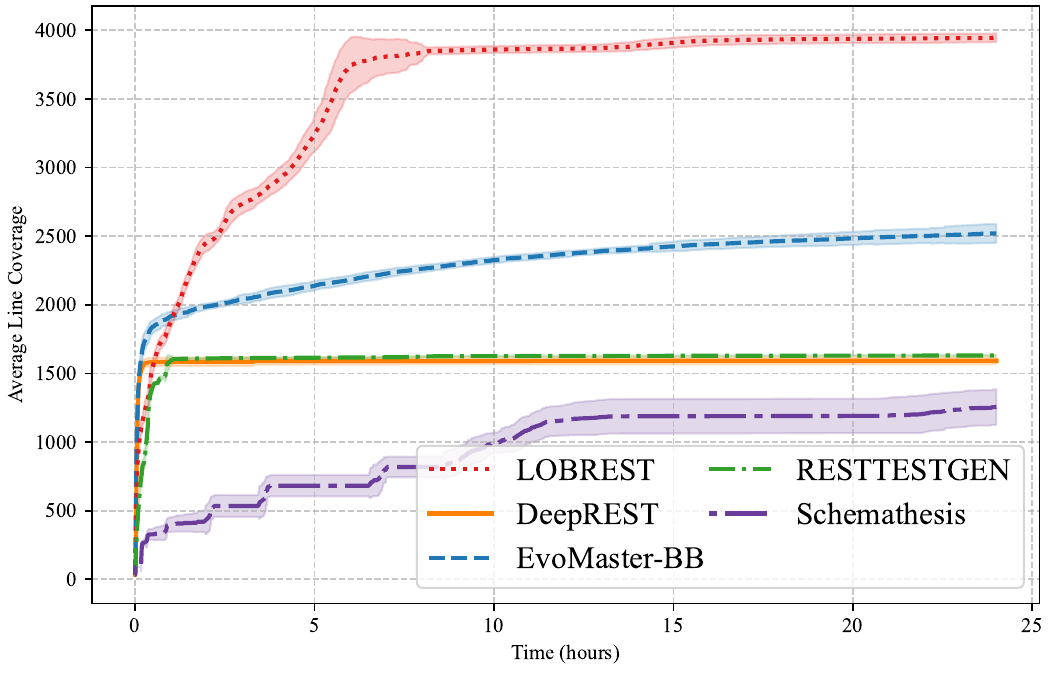} 
        \caption{Average line coverage on the whole GitLab over 24 hours.
        The lines represent the mean values across 20 independent runs, and the shaded areas indicate the standard deviation ($\pm$ SD).}
        \label{fig:cov_over_time}
    \end{minipage}
    \hfill 
    \begin{minipage}[b]{0.53\textwidth}
        \centering
        \includegraphics[height=4.3cm]{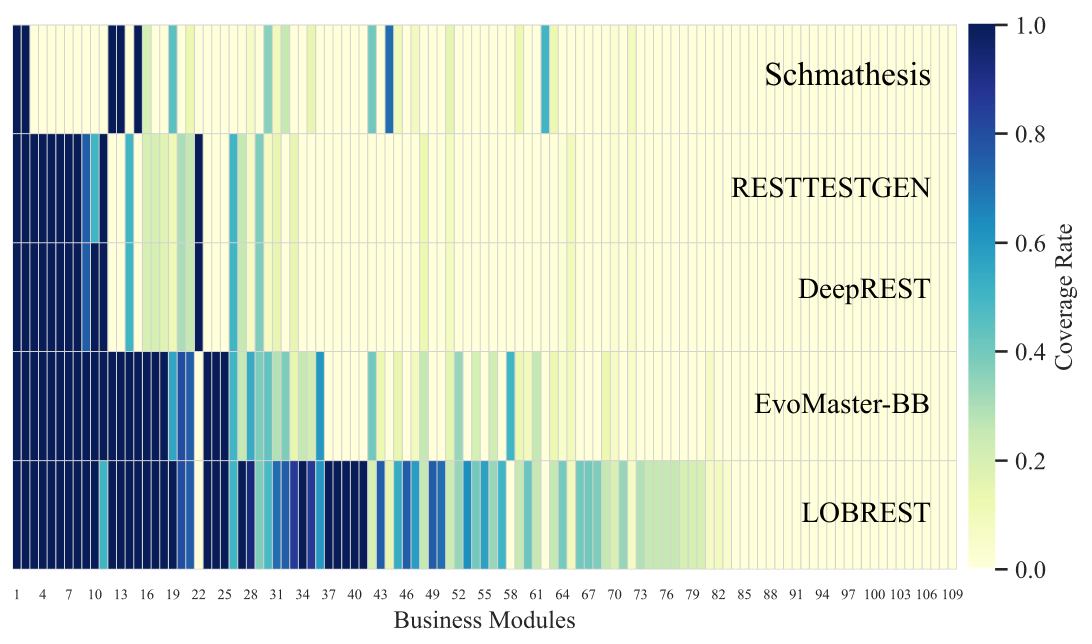} 
        \caption{Heatmap of coverage rates across different business modules for each testing tool.
        Each cell represents the coverage rate achieved by a tool for a specific module, with darker shades indicating higher coverage.}
        \label{fig:business_module_cov}
    \end{minipage}
    \Description{}
\end{figure}

\subsection{RQ2: Effectiveness on services with dense businesses}
GitLab is a large-scale, industrial service with 1,099 operations and 109 business modules \footnote{REST services are often organized into business modules, each grouping operations related to a specific business concern (e.g., Branch and Commit in GitLab). Such modules are developer-defined and can be reflected via operation tags.}, providing a wide range of business-sensitive functionalities.
Therefore, we categorize GitLab services S11-S17 as business services with dense businesses.
Evaluating \techname{} on such services demonstrates its capability to leverage HRLogs to recover business constraints, exercise complex functionalities, and achieve deep testing coverage in large-scale, real-world REST services.
Following prior studies \cite{atlidakis2019restler, wu2022restct}, we first test six commonly evaluated GitLab sub-services (S11-S16), with each run lasting one hour and repeated 20 times.
Next, we evaluate tools on the full GitLab service for the first time; considering the complexity of S17, each experiment on it lasts 24 hours and is repeated 20 times.

\subsubsection{Operation coverage.}
\autoref{tab:operation-coverage} presents the operation coverage achieved by \techname{} across S11-S17. \techname{} consistently attains the highest coverage, outperforming the strongest baseline, \restct{}, by 263.1\%.
Moreover, for the entire GitLab S17, \techname{} covers 352.7 operations, exceeding the second-best tool \evo{} by 188.6\%.
These results demonstrate that \techname{} exercises a substantially larger portion of the API operations compared to existing tools.

\subsubsection{Line Coverage.}
\autoref{tab:code-coverage} summarizes line coverage results on S11-S17.
\techname{} achieves the highest coverage on five sub-services, ranking second only on GitLab-Group (S13).
Averaged across the sub-services, \techname{} outperforms the strongest baseline \restct{} by 26.5\%.
On the entire GitLab S17, \techname{} exceeds the best-performing baseline \evo{} by 56.6\%.
\autoref{fig:cov_over_time} depicts the progression of average line coverage over 24 hours for \techname{} and four baselines.
Coverage growth for all tools plateaus during the latter half of the evaluation, indicating that the 24-hour time budget is sufficient for meaningful comparison.
Notably, \techname{} achieves superior coverage within the first hour and sustains this lead throughout the experiment, highlighting its efficiency in exercising code paths.
In summary, \techname{} demonstrates the strongest performance on line coverage, with particularly significant advantages when evaluated on the entire GitLab service.

\subsubsection{Business Module Coverage.}
\autoref{fig:business_module_cov} presents the coverage of GitLab business modules achieved by \techname{} and the four baseline tools during testing S17.
Each cell represents the coverage rate of a tool for a specific module, with darker shades indicating higher coverage.
\techname{} covers 80 out of 109 business modules, substantially surpassing the baselines: 55 for \evo{}, 32 for both \deeprest{} and \rtg{}, and 23 for \schemathesis{}.
In addition to breadth, \techname{} achieves the highest average operation coverage within modules at 45.4\%, outperforming the second-best tool, \evo{}, which reaches 27.9\%, by 62.7\%.
In summary, \techname{} excels in both the breadth and depth of business module coverage, demonstrating superior capability in exploring diverse business modules and effectively exercising their operations.

\subsubsection{Bug Detection.}
\autoref{fig:bugs}a and \autoref{fig:bugs}b present the bug detection results of \techname{} and the baseline tools on GitLab sub-services and the entire GitLab service.
On the sub-services, \techname{} detects 12 bugs, including 7 unique ones, outperforming all baselines.
The advantage becomes even more significant for the entire GitLab service: \techname{} detects 30 bugs—twice as many as the second-best tool \evo{}—including 21 bugs that other tools fail to find, exceeding the second-ranked \schemathesis{} by 133.0\%.
These results demonstrate that \techname{} is highly effective in detecting bugs in services with dense businesses compared to existing tools.

\begin{summarybox}
    \textbf{Answers to RQ2:}
    Compared with services with sparse businesses, \techname{} shows more pronounced advantages on services with dense businesses.
    Against the strongest baseline, \techname{} improves operation coverage by 263.1\% and 188.6\% and line coverage by 26.5\% and 56.6\% on S11–S16 and S17, respectively.
    \techname{} is also more business-aware, achieving 45.5\% and 62.7\% higher breadth and depth of business module coverage.
    Overall, \techname{} detects the most bugs; notably, on S17, it finds $2\times$ as many bugs as the second-best tool.
\end{summarybox}

\begin{figure}
    \centering
    \includegraphics[width=\linewidth]{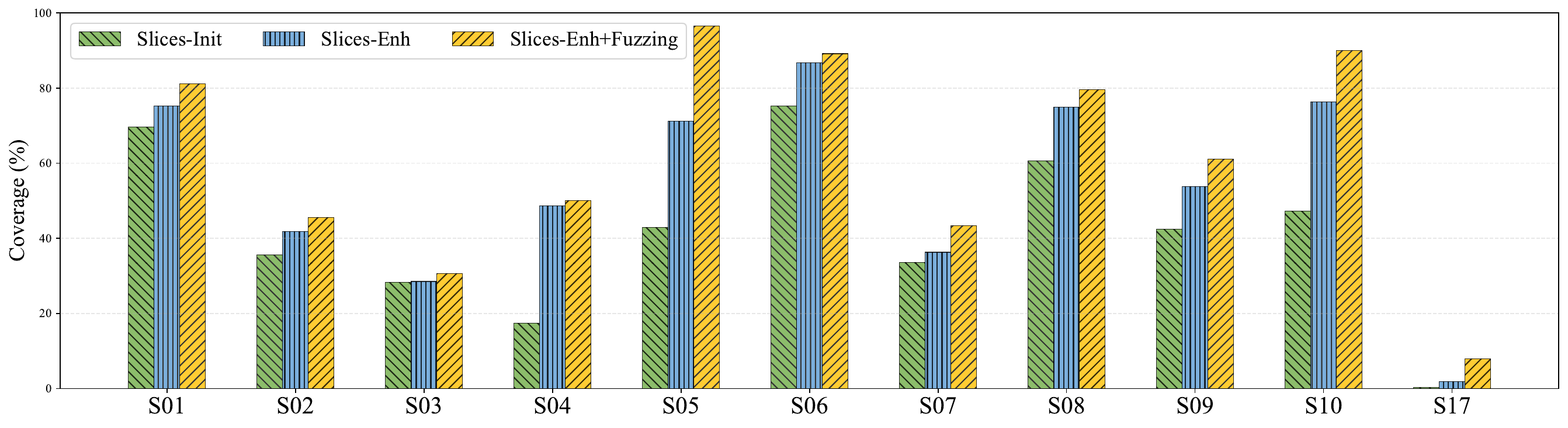}
    \caption{Line coverage comparison for initial slices, enhanced slices, and fuzzing with enhanced slices across 11 services.
    S11-S16 are excluded since their total line numbers are unknown.}
    \label{fig:ablation}
    \Description{Line coverage comparison for initial slices, enhanced slices, and fuzzing with enhanced slices across 11 services.
    S11-S16 are excluded since their total line numbers are unknown.}
\end{figure}

\subsection{RQ3: Contributions of Core Designs}

Log slicing, log slice enhancement, and business-aware fuzzing are key designs of \techname{} to repurpose HRLogs for REST API testing.
To evaluate their contributions to the effectiveness of LoBREST, we compare three configurations: the initial slice set $S$ (\emph{Slices-Init}), the enhanced slice set $S'$ (\emph{Slices-Enh}), and fuzzing with $S'$ (\emph{Slices-Enh+Fuzzing}).
\autoref{fig:ablation} shows the resulting line coverage for each configuration.
(1) The initial slice set achieves a certain level of coverage on all services, demonstrating that the locality-slicing strategy can effectively generate slices suitable for testing and provide crucial inputs for the subsequent stages.
(2) Across all 11 services, enhanced slices consistently achieve higher coverage than the initial slices, with an average improvement of 77.5\%.
This gain stems from \techname{} adding missing operations and completing required resource-creation entries within each slice.
(3) Using the enhanced slices as seeds for fuzzing further amplifies testing effectiveness, yielding an additional average improvement of 39.7\%.
Fuzzing is especially impactful for services with dense businesses: on GitLab, it increases the coverage of $S'$ by 313\%.

\begin{summarybox}
    \textbf{Answers to RQ3:}
    The core designs of \techname{} progressively improve testing performance.
    Initial log slices provide baseline coverage, generating operation sequences executable in testing.
    Enhancing the slices increases coverage by 77.5\% on average, and applying business-aware fuzzing on these slices further boosts coverage by 39.7\% on average, reaching up to 313.0\% on services with dense businesses like GitLab.
\end{summarybox}

\subsection{Threats to Validity}
\noindent
\textbf{Internal Validity.}
Our study is affected by three potential internal threats.
First, the effectiveness of the compared tools may be influenced by configuration choices, as different settings can lead to different testing behaviors.
To mitigate this threat and ensure fairness, we use the default configurations recommended in each tool's official repository.
Second, \techname{} relies on HRLogs, but real production logs of the evaluated services are unavailable.
To address this, we simulate realistic usage scenarios by recruiting participants to interact with the target services over a period of time, thereby generating HRLogs that approximate real-world usage patterns.
Third, the inherent randomness of fuzzing may affect the stability of the results \cite{klees2018evaluating, schloegel2024sok}.
To reduce this impact, we repeat the running of each tool on each service 20 times and report the average results.

\noindent
\textbf{External Validity.}
The threat to external validity arises from the limited number and diversity of the evaluated services \cite{ampatzoglou2019identifying}.
To mitigate this, we evaluate \techname{} on 17 REST services.
These services span multiple business domains and include both lightweight benchmarks and large-scale industrial systems.
Notably, we first evaluate testing tools on a complete GitLab service.
Such a large and real-world system provides indicative evidence of the general applicability of \techname{}.

\section{Related Work}
\textbf{REST API Testing.}
In recent years, various techniques have been proposed to ensure the reliability of REST services.
\restler{} \cite{atlidakis2019restler} is the first automatic stateful fuzzer, which generates test cases by inferring the producer-consumer dependencies among requests.
\evomaster{} \cite{arcuri2019evomaster} is another early REST API testing tool; it follows a white-box testing approach and applies evolutionary algorithms to optimize test case generation.
Wu et al. \cite{wu2022restct} proposed \restct{}, the first approach to apply combinatorial testing techniques to REST API testing.
Most existing works focus on generating operation sequences that satisfy resource constraints but overlook the importance of business constraints, resulting in insufficient exploration of the service.
Similar to us, \deeprest{} \cite{corradini2024deeprest} recognizes this issue and uses reinforcement learning to uncover implicit business logic and hidden constraints. However, \deeprest{} can only infer these constraints at runtime through repeated trial and error, because the specifications it relies on do not contain this information. In contrast, \techname{} leverages HRLogs, which inherently capture business constraints, allowing it to obtain constraint information more effectively and efficiently.


\noindent
\textbf{Log-Analysis in Testing.}
As an important means of profiling software systems, logging has become a common practice in enterprise operations. 
Many testing approaches leverage log data to enhance their testing effectiveness \cite{andrews1998testing, andrews2003general, aafer2021android}.
Wu et al. \cite{wu2024logos} present Logos, which transforms log data into semantic coverage information to guide the fuzzing process and enhance testing effectiveness.
Messaoudi et al. \cite{messaoudi2021log} proposed DS3, which shares a similar slicing-based approach with \techname{} for generating test cases.
Nevertheless, our approach differs greatly from DS3 in the slicing target, where DS3 slices complex system-level test cases with logs serving only as auxiliary guidance, whereas \techname{} directly slices logs to construct REST API test cases.


\section{Conclusion} \label{sec:conclusion}
This paper introduces \techname{}, a log-based, business-aware REST API testing technique that leverages historical request logs to facilitate effective testing of REST APIs.
With the locality slicing strategies, \techname{} first decomposes logs into smaller log slices, each slice preserving the required business constraints.
Then, the log slices are enhanced to ensure that the entire slice set can cover all API operations and each slice can be executed successfully.
Finally, the enhanced slices are used as initial seeds to perform business-aware REST API fuzzing.
We evaluate \techname{} against eight existing techniques on 17 REST services.
The results demonstrate that \techname{} consistently outperforms other techniques in terms of operation coverage, line coverage, and bug detection.

\bibliographystyle{acm}
\bibliography{a-bibfile}
\end{document}